\documentclass[a4paper,11pt]{article}
\pdfoutput=1 
\usepackage{soul}

\usepackage{jcappub} 
\usepackage[T1]{fontenc} 
\usepackage[normalem]{ulem}
\usepackage{float}

\title{On approximations of the redshift-space bispectrum and power spectrum multipoles covariance matrix}

\abstract{We investigate, in dark matter and galaxy mocks, the effects of approximating the galaxy power spectrum-bispectrum estimated covariance as a diagonal matrix, for an analysis that aligns with the specifications of recent and upcoming galaxy surveys. We find that, for a joint power spectrum and bispectrum data-vector, with corresponding $k$-ranges of $0.02<k\,[h{\rm Mpc}^{-1}]<0.15$ and $0.02<k\,[h{\rm Mpc}^{-1}]<0.12$ each, the diagonal covariance approximation recovers $\sim 10\%$ larger error-bars on the parameters $\{\sigma_8,f,\alpha_\parallel,\alpha_\bot\}$ with respect to the full covariance case, while still underestimating the corresponding true errors on the  recovered parameters by $\sim 10\%$. This is caused by the diagonal approximations weighting the elements of the data-vector in a sub-optimal way, resulting in a less efficient estimator, with poor coverage properties,  than the maximum likelihood estimator featuring the full covariance matrix. We further investigate intermediate approximations to the full covariance matrix, with up to $\sim 80\%$ of the matrix elements being zero, which could be advantageous for theoretical and hybrid approaches. We expect these results to be qualitatively insensitive to variations of the total cosmological volume, depending primarily on the bin size and shot-noise, thus making them particularly significant 
for present and future galaxy surveys.}

\author[1]{Sergi Novell-Masot,}
\author[1]{H\'ector Gil-Mar\'in}
\author[1,2]{and Licia Verde}
\affiliation[1]{ICC, University of Barcelona, IEEC-UB, Mart\' i i Franqu\` es, 1, E08028
Barcelona, Spain}
\affiliation[2]{ICREA, Pg. Lluis Companys 23, Barcelona, 08010, Spain} 
\emailAdd{sergi.novell@icc.ub.edu, hectorgil@icc.ub.edu, liciaverde@icc.ub.edu}

\begin{document}
\maketitle

\section{Introduction}

The foundation of cosmological parameters inference rests on the   fact  that the maximum likelihood estimator (MLE) is the best unbiased estimator in large data sets (e.g.,\cite{tegmark1997karhunen} for a cosmologist-friendly presentation). In most applications, and in particular in  large-scale structure clustering, the likelihood is assumed to be Gaussian. 

In this context, while the power spectrum is the primary large-scale structure clustering statistic, it has been shown that including the redshift space bispectrum significantly enhances the constraining power of galaxy  surveys (e.g., \cite{2dFBispectrum,ScoccimarroIRAS,HGMbispectrum15b}), by breaking cosmological parameter degeneracies and reducing error-bars on key cosmological parameters.

In principle, even if the initial density field were to be Gaussian, the distribution of the $n$-point functions, such as the three-point correlation function (3PCF) or the bispectrum (its counterpart in Fourier space) is not Gaussian. However, these summary statistics are usually binned (in $k$ or scales): for large survey volumes, each bin is populated by many data points and, thanks to the central limit theorem, the resulting statistics for the bins approach Gaussianity. In particular, the joint likelihood of the power spectrum and bispectrum multipoles is assumed to be a multi-variate Gaussian, and this approximation has been shown to hold well in practice,  in regimes without too strong non-linearities \cite{carron2013assumption,dodelson2013effect,repp2015impact,bellomo_beware_2020,Gualdi:2020aniso}. 

Once a data-vector and a model for the data-vector (also referred to as a signal) are given, the full covariance  matrix is the key ingredient for cosmological inference.
The evaluation of this matrix beyond the purely linear regime  is challenging: analytic expressions 
quickly become long and cumbersome to evaluate \cite{biagetti2022covariance,Gualdi:2020aniso} especially if all real-world effects present in a real survey need to be taken into account. 
Alternatively, the covariance matrix can be evaluated from multiple mock survey realizations  (see for example \cite{novell2023geofpt,oddo2021cosmological,gualdi_matter_2021,gualdi_joint_2021} for the specific case involving the redshift space bispectrum multipoles). In this case, the number of realizations needs to be  significantly larger than the number of data-vector elements \cite{Sellentin:2015waz,Hartlap:2006kj,percival2014clustering,percival2022matching}. Here the challenge lies in the fact that when including bispectrum multipoles the data-vector can easily reach a length of $\sim 1000$ elements, requiring  sometimes a prohibitive number of mocks. 

All these challenges could be significantly eased if the  off-diagonal terms of the covariance could be ignored. Some work in the literature has advocated for this approach. For instance  \cite{Gil-Marin:2014biasgravity} used a diagonal covariance with diagonal elements estimated from a set of simulations. 
A different "flavour" of diagonal covariance matrix is the so-called  Gaussian covariance matrix  which is not only diagonal but also non-linearities are neglected. Several studies have been using  a Gaussian covariance matrix: either modelling it perturbatively \cite{ivanov2022precision,ivanov2023cosmology} or using a Gaussian random field template (e.g., \cite{slepian2015computing,slepian2017detection,slepian2017large} for the 3-point correlation function, and \cite{xu20122} for the 2-point correlation function). In these (non-exhaustive) references, the choice of Gaussian covariance is justified by invoking the linearity of the scales present in the analysis.
In summary, the motivation behind the choice of a diagonal covariance matrix is not only the simplicity but also that the resulting (forecasted) errors  do not seem to be much affected. It would not be remiss to hope that this could be a sufficiently good approximation in the quasi-linear regimes, where all these analyses are based.

For present and upcoming galaxy clustering analyses, such as DESI \cite{aghamousa2016desi}, Euclid \cite{laureijs2011euclid} or LSST \cite{ivezic2019lsst}, relevant information  will be extracted from scales where non-gaussianities are not negligible. Notable efforts in developing analytical templates for higher-order correlators with non-Gaussian contributions have been developed by e.g. \cite{sugiyama2020perturbation,sugiyama2019complete,Gualdi:2020aniso}. It is then worthwhile and timely to investigate the magnitude of the penalty incurred when using a diagonal approximation for a bispectrum analysis up to mildly non-linear scales.

In the context of a MLE for cosmological inference, using a diagonal covariance matrix would imply  adopting an estimator that,  while being still (asymptotically) unbiased, is not the best-unbiased estimator---see e.g. \cite{floss2023primordial} for primordial non-Gaussianities studies. In other words, the weighting given to the data-vector elements is not optimal. 
In this paper, we  elucidate the importance of including the off-diagonal elements in the covariance matrix 
for cosmological inference from the bispectrum summary statistic.

The rest of this paper is organized as follows. In Section~\ref{sec: sims} we present the suites of simulations we use and the set-up. In Section \ref{sec: results} we compare the effects of assuming a diagonal covariance matrix, and we review a few intermediate alternative approximations in Section \ref{sec: approxs}. In Section~\ref{sec: theory} we offer an analytic description of the effects found on simulations. Finally, we conclude in Section~\ref{sec: concl}.

\section{Simulations and set-up}
\label{sec: sims}
We review the main aspects of our analysis set-up and simulation data, which is very similar to what we used in \cite{novell2023geofpt}---to which we refer the readers for a more detailed description. The two main  simulation sets used in this work are the \textsc{Quijote} suite \cite{villaescusa-navarro_quijote_2020} and the \textsc{Patchy} galaxy mocks \cite{kitaura2016clustering}. These are cubic periodic boxes where real-world effects such as window function, selection function, or systematic weights, are not included. The volume of the \textsc{Quijote} realizations is 1 $(\textrm{Gpc}\,h^{-1})^3$, whereas the \textsc{Patchy} galaxy mocks have volume 15.6 $(\textrm{Gpc}\,h^{-1})^3$ each.

In both cases, the data-vector consists of the power spectrum and bispectrum monopole and quadrupoles: ${\cal D}=\{P_0,P_2,B_0,B_{200},B_{020}\}$. We use the bispectrum multipoles expansion first derived in \cite{Scoccimarro:1999ed}. Note that we have neglected one bispectrum quadrupole configuration, $B_{002}$, as it has been shown that it does not add information to our analysis in the scales of interest \cite{d2022boss,novell2023geofpt}. We discard the isosceles configurations of the form $(k_1,k_1,k_2)$ of the $B_{020}$ data-vector, due to them being redundant: $B_{200}(k_1,k_1,k_2)=B_{020}(k_1,k_1,k_2)$ for all possible $k_1,k_2.$

We consider $k$-ranges of $0.02<k^{P}\,[h{\rm Mpc}^{-1}]<0.15$ and $0.02<k^{B}\,[h{\rm Mpc}^{-1}]<0.12$, with the superscripts $P$ and $B$ indicating power spectrum and bispectrum respectively. Following Ref.~\cite{gualdi_joint_2021} the $k$-vectors are binned  with $\Delta k=1.1k_{\rm f}\approx 0.0069\,h\,{\rm Mpc}^{-1}$ (with $k_{\rm f}=2\pi/L_{\rm box}$ the fundamental frequency) for \textsc{Quijote}, while for \textsc{Patchy} the binning is such that $\Delta k=0.01\,h\,{\rm Mpc}^{-1}$. 

We model the power spectrum up to two loops in renormalized perturbation theory (RPT) \cite{Crocce_2006,Gil_Mar_n_2012_pert}, whereas for the bispectrum the recently presented phenomenological model GEO-FPT\footnote{\url{https://github.com/serginovell/geo-fpt}} \cite{novell2023geofpt} is employed. The full parameter space is given by $\{\sigma_8,\,f,\,\alpha_\parallel,\,\alpha_\bot,\,b_1,\,b_2,\,A_{P},\,\\\sigma_{ P},\,A_{B},\, \sigma_{B}\}$. These are, respectively, the amplitude of dark matter fluctuations smoothed by a top-hat filter of $8\,{\rm Mpc}h^{-1}$,  $\sigma_8$, the logarithmic growth rate, $f$, the dilation parameters along and across the line-of-sight, $\alpha_\parallel$ and $\alpha_\bot$ \cite{Alcock:1979mp,ballinger1996measuring,beutler2014clustering,gil-marin_clustering_2017}; the linear and non-linear galaxy bias parameters $b_1$, $b_2$, where the local Lagrangian bias expansion is assumed \cite{baldauf2012evidence,saito2014understanding,Brieden_ptchallenge};\footnote{This expansion fully determines the tidal bias and non-linear bias, respectively $b_{s^2},b_{3nl}$, from the linear bias parameter $b_1$.} the phenomenological amplitude parameters that regulate the deviations from shot-noise ($A_{P},A_B$ \cite{Gil-Marin:2014biasgravity,gil-marin_clustering_2017}), and the Fingers-of-God damping factors $\sigma_{ P},\sigma_{ B}$ \cite{Jackson_1972}, considered independent for the power spectrum and bispectrum. The subset $\{\sigma_8,f,\alpha_\parallel,\alpha_\bot\}$  holds more relevance for cosmology and is the focus of this work, while the rest are treated  as nuisance parameters.

We use 8000 realizations of the fiducial set of \textsc{Quijote} dark matter simulations at redshift $z=0.5$ to estimate the full covariance matrix. This large number of realizations ensures a reliable estimate of the covariance, for a fairly large number of elements in our data-vector ($\sim 1200$). Then, we also consider 2000 (of these 8000) simulations to be analyzed individually, thus obtaining 2000 sets of (single realization) best-fit parameters. We compare the distribution of these best-fit parameters with the MCMC posteriors derived by fitting the averaged data vector of the same 2000 simulations.

Likewise, for the \textsc{Patchy} mocks we employ the available 2000 realizations  at redshift $z=0.53$; we fit cosmological parameters from  them individually, and from their  mean  power spectrum and bispectrum signal (to reduce cosmic variance). In this case, the  full covariance matrix is estimated from 2000 realizations, as this is the maximum available number.

\begin{figure}[h!]
\centering 
\includegraphics[width = 0.8\textwidth]{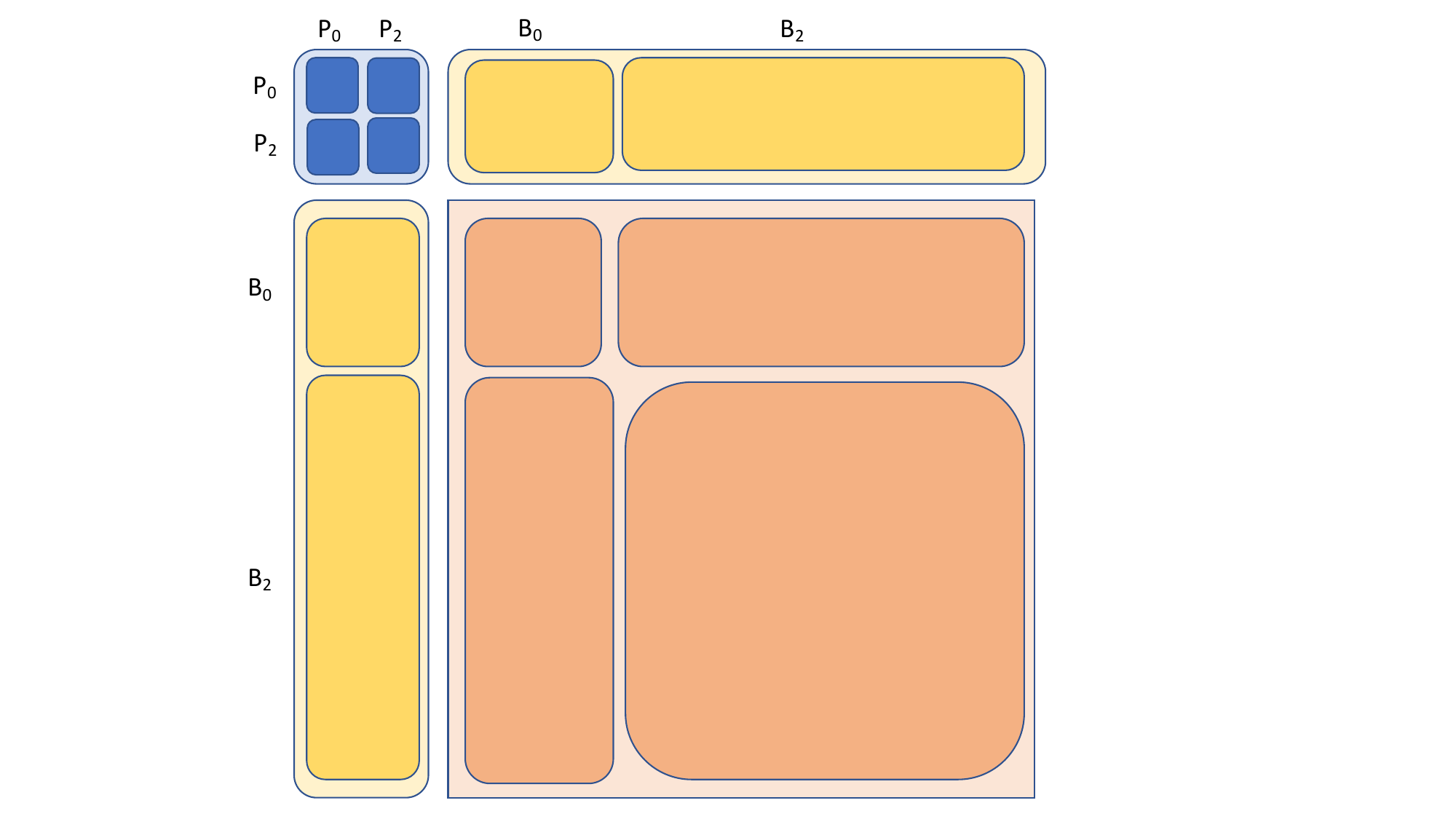}
\caption{Schematic representation of the full power spectrum-bispectrum covariance matrix. Here $P_0$  denotes the power spectrum monopole portion of the  data-vector, $P_2$ being the power spectrum quadrupole portion. Similarly, $B_0$ and $B_2$ indicate the bispectrum monopole and quadrupole portions of the data-vector respectively. In the {\it full} approximation the matrix is dense in the sense that all elements  of all blocks are non-zero even though many are small.  In the {\it diagonal} approximation the off-diagonal blocks are all zero  and the blocks on the  diagonal, %
 e.g., the $P_0-P_0$, $B_0-B_0$ and $B_2-B_2$ blocks, are diagonal matrices. Intermediate approximations discussed in Section \ref{sec: approxs} populate the off-diagonal block and the off-diagonal elements of the diagonal blocks according to some  case-specific rules.  The blue, yellow, and orange sub-matrices will be referred to as $PP$, $PB$, and $BB$ blocks, respectively.}
\label{fig: sketch}
\end{figure}

The full power spectrum and  bispectrum covariance matrix can be schematically represented in blocks as shown  in Figure~\ref{fig: sketch}. In all cases, at the likelihood evaluation level,  we correct for the inverse covariance matrix estimation bias using the Sellentin and Heaven prescription \cite{Sellentin:2015waz}. To estimate the magnitude of this effect it is useful to report that the length of the data-vector is 1160 for \textsc{Quijote} and 441 for \textsc{Patchy}; given the available number of simulations, the corresponding Hartlap correction factor is respectively $H=0.85$ for  \textsc{Quijote},  and  $H=0.78$ for \textsc{Patchy}.\footnote{While the correction is performed using the Sellentin-Heavens prescription,  the Hartlap factor (see Equation 17 of \cite{Hartlap:2006kj}) $H=(n-m-2)/(n-1)$ where $n$ is the number of simulations and $m$ the size of the data-vector, only indicates an order of magnitude of the noise introduced  when inverting the covariance.}

\section{Diagonal approximation: effects in simulations}
\label{sec: results}

In order to quantify the effect of assuming a diagonal covariance matrix, we compare the cosmological constraints for an analysis involving the power spectrum and bispectrum monopole and quadrupoles, using both the full covariance matrix and the diagonal approximation. In this approximation,  all off-diagonal correlations for the power spectrum, the bispectrum, and the cross power-spectrum bispectrum are set to zero (see Figure~\ref{fig: sketch}). 

In what follows we compute constraints (i.e., recovered values and estimated errors) for the key cosmological parameters under different approximations and conditions.
For both dark matter N-body (\textsc{Quijote}) and galaxy (\textsc{Patchy}) simulations we compute the four cases that are specified in Table \ref{tab:comparison_methods}.

\begin{table}[ht]
\centering
\begin{tabular}{|p{3.2cm}||p{5.5cm}|p{5.5cm}|}
\hline

& \textbf{Posterior of the mean data vector} & \textbf{Distribution of individual best-fits} \\ \hline\hline
\textbf{Full Covariance Matrix} & Posteriors, recovered via MCMC using full covariance matrix, from the mean of the data vectors of  2000 simulations  & Distribution of   best fit parameters, recovered via MCMC using full covariance matrix,  for the 2000 individual simulations \\ \hline
\textbf{Diagonal Covariance Matrix} & Posteriors, recovered via MCMC using diagonal covariance matrix, from the mean of the data vectors of 2000 simulations & Distribution of   best fit parameters, recovered via MCMC using diagonal covariance matrix,  for the 2000 individual simulations \\ \hline
\end{tabular}
\caption{The four methods of inference we perform on both \textsc{Patchy} and \textsc{Quijote} simulations. For respectively the full covariance and diagonal approximation, we perform both the standard MCMC sampling for the parameters of interest, and the distribution of best-fit parameters obtained individually in each of the 2000 simulations.  }
\label{tab:comparison_methods}
\end{table}

Any non-optimal weighting is expected to produce  uncertainties on the inferred parameters that are larger than those achievable with an optimal weighting. 
Moreover, the  inferred posterior provided by the MCMC  might  not yield error-bars that have the correct coverage properties \cite{sellentin2019debiasing}.\footnote{The statistical concept of coverage refers to the probability that a confidence interval correctly contains the true values of the parameters being estimated. These confidence intervals do not necessarily coincide with the error-bars derived via MCMC---the \textit{credible intervals}---, which represent the probability range of parameters based on prior beliefs and observed data.} On the other hand, the scatter among individual simulations' best-fit parameters is expected to have better coverage properties for a large enough number of simulations.

Throughout the following subsections, the diagonal covariance approximation is confirmed to  underestimate the errors of the recovered  cosmological parameters for analysis set-ups similar to the one in  Section~\ref{sec: sims}. Section \ref{sec: approxs} tackles the intermediate case where not enough simulations are available in order to estimate or calibrate the full covariance matrix. There, other  sparse matrix approximations  to the full covariance matrix  are then offered, along with their potential for  reducing the number of required simulations for covariance estimation.

\begin{figure}[htb]
\centering 
\includegraphics[ width = 0.49\textwidth ]
{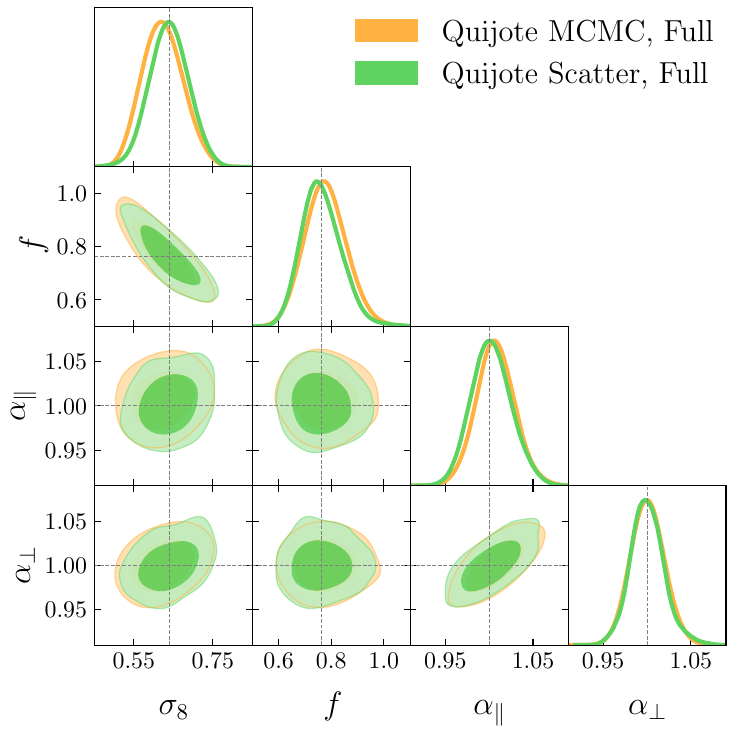}
\includegraphics[width = 0.49\textwidth ]
{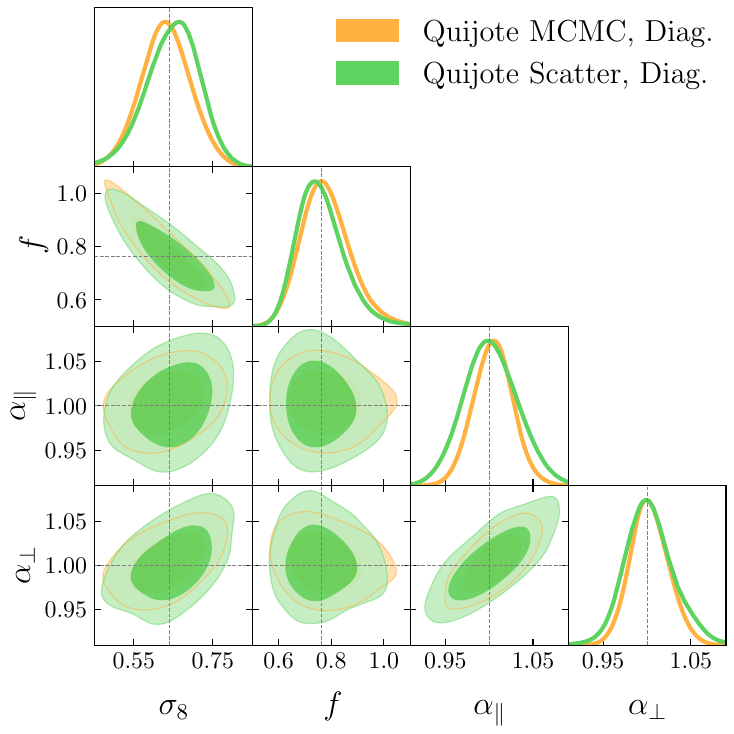}
\includegraphics[ width = 0.49\textwidth ]
{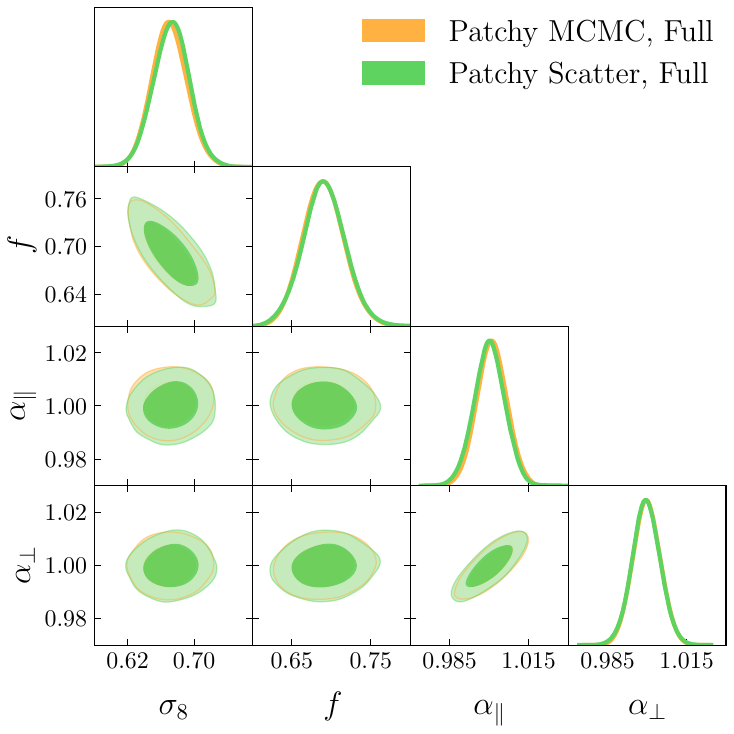}
\includegraphics[width = 0.49\textwidth ]
{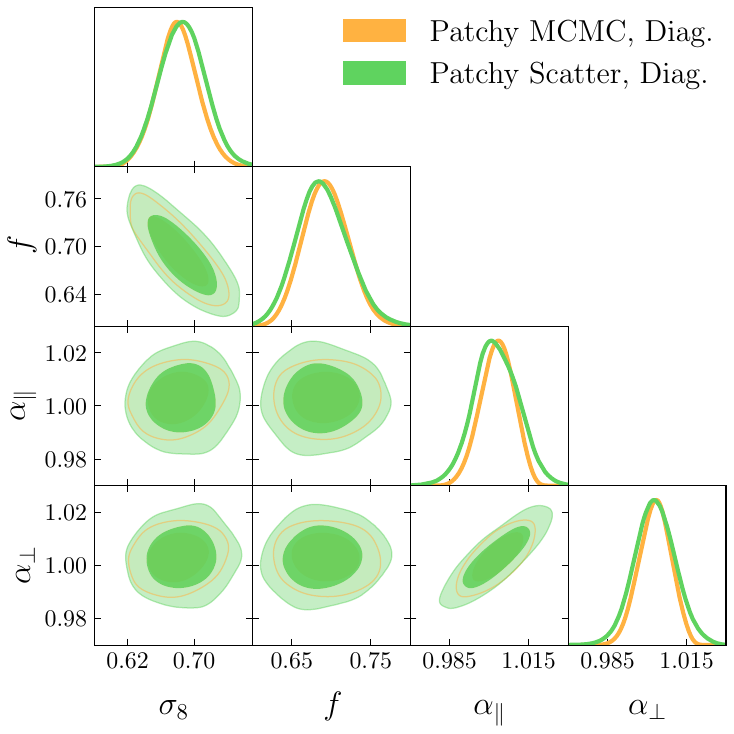}

\caption{Estimated posteriors for the main cosmological parameters with the data-vector $P_0+P_2+B_0+B_{200}+B_{020}$ from 2000 \textsc{Quijote} fiducial dark matter simulations (upper row) and the 2000 \textsc{Patchy} galaxy FastMocks (lower row). In the left plots, the full covariance (estimated from the simulations) is employed,  while in the right plots, the diagonal approximation is used. The orange contours are {\it credible intervals} estimated from the  posteriors sampled via MCMC, using the mean of the 2000 simulations as a data-vector. The green contours are obtained from the scatter of the maximum likelihood parameters for  each of the 2000  simulations individually, hence represent {\it confidence intervals}. The distribution inferred from the scatter of the maximum likelihood values is smoothed via a kernel density estimator to provide smooth confidence contours to compare with those inferred from the MCMC. The power spectrum is modeled at 2L-RPT and for the bispectrum we employ the GEO-FPT model presented in \cite{novell2023geofpt}. The axes size for each set of simulations is kept fixed to facilitate visual comparison of the posteriors between the two covariance approaches.
Using the diagonal approximation produces {\it a)} larger scatter hence larger actual uncertainties,  and {\it b)} causes the  MCMC approach to  underestimate the (actual) errors.}
\label{fig: mcmc_vs_scatter}
\end{figure}

\subsection{Parameter constraints: MCMC and scatter}

Given 2000 independent realizations of both the \textsc{Quijote} dark matter N-body simulations and the \textsc{Patchy} galaxy mocks (presented in Section \ref{sec: sims}), we proceed to compare the recovered parameters $\{\sigma_8,f,\alpha_\parallel,\alpha_\bot\}$ and their estimated uncertainties with two different methods. In particular, we contrast the constraints obtained via the traditional approach of MCMC sampling of the posterior distributions (credible intervals), with the scatter of maximum likelihood values for each realization (confidence intervals). If the statistical approximations adopted---mainly the Gaussian likelihood and optionally the diagonal covariance---are correct, then  the MCMC posteriors would have correct coverage properties and thus the two approaches would yield comparable error-bars.

 Figure \ref{fig: mcmc_vs_scatter}  shows that in both sets of simulations, the diagonal covariance MCMC fails to capture the information that is obtained with the scatter of maximum likelihood values, especially for the dilation parameters $\alpha_\parallel,\alpha_\bot$. 
 The diagonal approximation produces {\it a)} a larger scatter hence larger actual uncertainties; and {\it b)} an incorrect  posterior estimate via the MCMC, which produces an underestimate of the actual errors (i.e., the diagonal approximation yields credible intervals that are smaller than the corresponding confidence intervals and thus have no {\it coverage}). 
 This behavior is not unexpected at all: the diagonal covariance approximation makes the adopted estimator and likelihood suboptimal as anticipated above and discussed also in \cite{sellentin2019debiasing,percival2022matching}. 
 
 In the \textsc{Quijote} simulations, the scatter of maximum likelihood values in the diagonal approximation is noticeably larger than the MCMC posteriors, while
 in the \textsc{Patchy} mocks the effect is milder, mostly affecting the tails of the distribution. 
 This may be caused by the combination of the difference in tracer type, shot-noise and bin-size; in particular, \textsc{Quijote} has negligible shot-noise and a smaller bin size, as specified in Section \ref{sec: sims} and in \cite{novell2023geofpt}. For simplicity, 
 in what follows we focus on the \textsc{Patchy} mocks.\footnote{We have checked that the results presented hereafter are qualitatively equivalent in both \textsc{Quijote} and \textsc{Patchy} simulation sets, with a small difference in magnitude. We choose to present the results from the \textsc{Patchy} mocks for two  reasons: firstly, their closer resemblance to galaxy data compared to the \textsc{Quijote} simulations; and secondly, as demonstrated in Figure \ref{fig: mcmc_vs_scatter}, the \textsc{Patchy} simulations exhibit superior performance under the diagonal approximation. This implies that, should the diagonal approximation prove unsuitable for \textsc{Patchy}, its performance in the \textsc{Quijote} simulations would be equally, if not more, limited.}

\subsection{Full and diagonal covariance: effects on $\alpha_\parallel$ and $\alpha_\perp$}

Motivated by the fact that the effect of the diagonal approximation seems to propagate mainly to the dilation parameters, in the main text we only show and discuss  the results for $\{\alpha_\parallel,\alpha_\bot\}$. The  $\sigma_8$ and $f$  cases are displayed in Appendix \ref{app: fs8}.\footnote{This difference between the two sets of parameters could be due to there being more proportion of information content for $\alpha_\parallel,\alpha_\bot$ in smaller scales (where the diagonal approximation performs the worst) than for $f,\sigma_8$.}

\begin{figure}[htb]
\centering 
\includegraphics[ width = \textwidth ]
{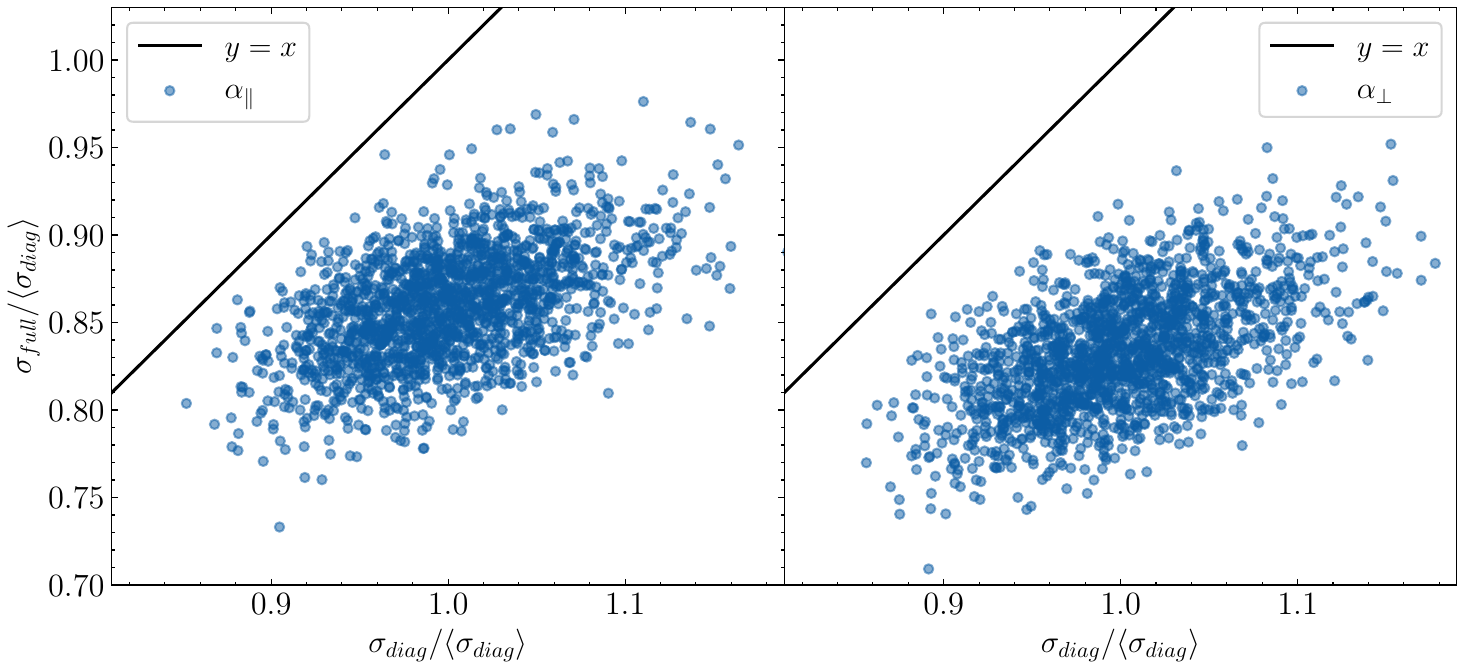}
\caption{Scatter of the marginalized 1$\sigma$ regions  for $\alpha_\parallel$ (left) and $\alpha_\bot$ (right) from 2000 \textsc{Patchy} mocks. 
On the  $x$-axis we show  the MCMC-recovered 1$\sigma$ region using the diagonal covariance, $\sigma_{\rm diag}$, on the  $y$-axis the  corresponding 1$\sigma$ region, using the full covariance matrix, $\sigma_{\rm full}$. Both are normalized by the mean 1$\sigma$ region with the diagonal covariance matrix, $\langle\sigma_{\rm diag}\rangle$ for easy visualization. Each point corresponds to one of the 2000 \textsc{Patchy} realizations.
 The identity line $y=x$ is over-plotted, together with the best-fit regression line to the scatter of points. The recovered error-bars from  MCMCs  are larger when using a diagonal covariance matrix ($\sigma_{\rm diag}\sim 1.2\sigma_{\rm full}$).} 
 
\label{fig: errors_corr_alpha}
\end{figure}

Figure \ref{fig: errors_corr_alpha} presents the relationship between the recovered error-bars, obtained via MCMC,  using a diagonal ($\sigma_{\rm diag}$) and full ($\sigma_{\rm full}$) covariance, for the dilation parameters. 
The identity line is also shown for guidance.
As expected from Figure~\ref{fig: mcmc_vs_scatter} the diagonal approximation yields larger error-bars ($\sigma_{\rm diag}\sim1.2\sigma_{\rm full}$). 
In addition, a weak correlation is observed between the derived errors from the full covariance and from the diagonal part only.

An alternative perspective on the impact of the diagonal covariance matrix approximation can be gained as follows. Let us consider the statistic $z$, defined as \cite{BAO_RSD},
\begin{equation}
\label{eq: def_z}
    z_{\theta_j}\equiv\frac{\theta_j-\mu_{\theta_j}}{\sigma(\theta_j)},
\end{equation}
where $\theta_j$ denotes one of the parameters of interest, i.e. $\theta_j\in\{\sigma_8,f,\alpha_\parallel,\alpha_\bot\}$. For each $i$-realization of the 2000 simulations, $\theta_j^i$  denotes the best-fitting parameter value recovered via MCMC and $\sigma(\theta_j^i)$ is its error estimated from the MCMC posterior; $\mu_{\theta_j}$ denotes the mean  across all simulations of the recovered parameter $\theta_j^i$. We expand on the features of the $z$ statistic in Appendix \ref{app: z}.

Since the simulations are effectively independent and identically distributed realizations of the underlying fiducial cosmology, if the errors are correctly estimated, the distribution of the 2000 values of $z_{\theta_j}$, by the central limit theorem, should converge to a normal with variance 1, $\mathcal{N}(0,1)$. Consequently, $\sigma_z$ (the standard deviation of the $z$ statistic across the 2000 realizations) should converge to 1. Some (small) mis-match is expected, for at least  two reasons. The estimated covariance, which enters in the denominator of Equation~\ref{eq: def_z} has an associated error, which means that the true covariance could have been different than that adopted \cite{kodwani2018effect,Hartlap:2006kj}. The likelihood is non-Gaussian so matching the {\it rms}  with the 1-$\sigma$ of a Gaussian distribution may not be exact.  We return to the first effect below; nevertheless relative statements can still be made. 

\begin{figure}[htb]
\centering 
\includegraphics[ width = \textwidth ]
{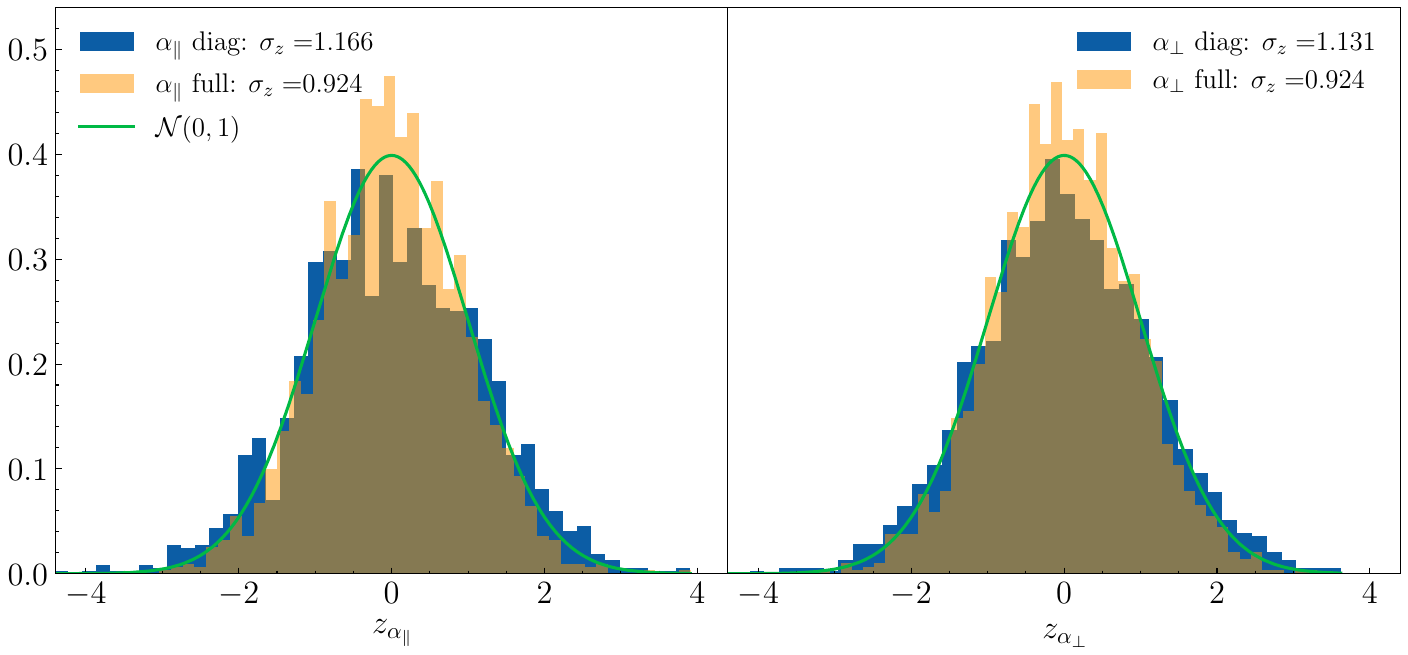}

\caption{Histogram of the $z$ values for the full and diagonal covariance matrix cases, for the parameters $\alpha_\parallel$ (left panel) and $\alpha_\bot$ (right panel). The standard deviation of $z$ ($\sigma_z$) is shown for each case. The solid line, the standard normal distribution $\mathcal{N}(0,1)$, corresponds to  the theoretical shape of $z$ where the errors $\sigma(\theta)$ are perfectly estimated. 
The diagonal approximation (blue) tends to underestimate the error-bars in the cosmological parameters, while the full covariance (orange) instead tends to overestimate them. }
\label{fig: histz_alpha}
\end{figure}
Figure \ref{fig: histz_alpha} shows the distribution of $z_{\alpha_\parallel}$ (left panel) and $z_{\alpha_\perp}$ (right panel) for  the full (orange) and diagonal (blue) covariance matrix cases. 
The deviations from  $\mathcal{N}(0,1)$ for $z_{\theta}$  offer a metric to assess  the accuracy in the estimation of the errors. From the  definition of $z_{\theta}$ in Equation \ref{eq: def_z}, if the error-bars on the cosmological parameters are consistently underestimated (resp. overestimated),  then $\sigma_z>1$ (resp. $\sigma_z<1$)---with $\sigma_z$ being the standard deviation for the distribution of $z_{\theta}$. For example, for the two dilation parameters we find that  for the full covariance $\sigma_z=0.924$ while for the diagonal approximation $\sigma_z=1.17\, (1.13)$ for $\alpha_\parallel$($\alpha_\perp$). Hence the true errors appear  to be over-estimated by about $\sim8\%$ when employing the full covariance;  when employing the diagonal matrix they appear under-estimated by $17\%$($13\%$) for $\alpha_\parallel$($\alpha_\perp$).

This figure confirms  the findings of  Figure \ref{fig: mcmc_vs_scatter}:  the cosmological inference results  obtained  from running an MCMC assuming a diagonal covariance 
will underestimate the error-bars,  and the constraints will also be sub-optimal. Nevertheless, we find no evidence of a systematic shift in the central values of the recovered parameters: the recovered best fit  parameters are unbiased.  
This again boils down to the diagonal approximation not being a maximum likelihood estimator---the estimator is unbiased but sub-optimal.
The inference using the full covariance matrix as estimated from the 2000 simulations and the Sellentin and Heavens correction \cite{Sellentin:2015waz} appears to be, according to the $z_\theta$ metric,  slightly conservative. 

We repeated the same procedure  also for the  \textsc{Quijote} set ($H=0.85$) and for the power spectrum part of the signal, $P_{02}$, in the  \textsc{Patchy} set (the reduced size of the data vector yields $H=0.99$). In these cases we find  $\sigma_z=1$ (\textsc{Patchy}, $P_{02}$ only) and $\sigma_z=0.97$ (\textsc{Quijote}, $P_{02}+B_{02}$) for the $\alpha$ parameters, i.e. the errors are  correctly estimated at better than 3\%.   As expected, the parameters which have markedly non-Gaussian posterior tails (i.e. $f, \sigma_8$ from the power spectrum alone), do not perform as well.
For comparison  diagonal covariance matrices  mis-estimate the error by $\sim 20\%$.
We tentatively conclude that the appearance of $\sigma_z$ slightly above unity ($\sim  8\%$) for the full covariance case in Figure \ref{fig: histz_alpha} is correlated to  the noise in the estimation of the covariance matrix, which is driven by  the  number of simulations and the size of the data vector as quantified by the Hartlap factor.

We will not dwell more on this except for drawing the following  conclusion:   while  the full covariance matrix estimated from the available number of simulations is not the ground truth, it is good enough to be treated as our baseline for comparison for various approximations. The diagonal approximation  underestimates the uncertainties by $\sim 15\%$.

\section{Intermediate approximations}
\label{sec: approxs}
The diagonal approximation for the power spectrum and bispectrum covariance matrix may therefore be a good approximation for Fisher-matrix-based  error forecasts, but not accurate enough for precision cosmology from ongoing surveys.
We may however investigate whether there are sparse matrix approximations to the full covariance that may potentially require fewer simulations to estimate \cite{bickel2008covariance,cai2011adaptive,cai2012optimal,friedman2008sparse,oztoprak2012newton,padmanabhan2016estimating} and perform  effectively as well as the full covariance.\footnote{Note that in the last reference, \cite{padmanabhan2016estimating}, which holds special interest given its formulation in terms of cosmology, the focus is on the sparsity of the precision matrix, the inverse of the covariance matrix.  }  We reiterate  that the full covariance matrix estimated from simulations is not the ground truth but it is the best estimate we have which therefore we use as our baseline.
There is significant literature associated with modeling with random matrices the properties of noisy estimates of covariance matrices. We will only refer to this tangentially  here.

In what follows, we explore a small set of possible approximations, leaving out of the scope of this work valuable contributions that go towards the same direction, such as data-vector compression \cite{heavens2017massive,gualdi_geometrical_2019,Gualdi:2019sfc}, covariance matrix shrinkage \cite{stein1972improving,taylor2013putting,joachimi2017non} or resampling \cite{efron1982jackknife,schneider2011fast}.  

We refer to $PP$, $PB$, and $BB$, respectively, as the  blue,  yellow, and orange  sub-matrices of the full matrix shown in Figure~\ref{fig: sketch}. 
We consider the following approximations, all of them  being positive-definite, having the full power spectrum auto-covariance $PP$ and of course all the diagonal terms:
\begin{itemize}
    \item ``Auto'': Include the full $PP$ and $BB$ boxes, setting to zero the $PB$ cross-covariance.
    \item ``Common-$k$'': Include the full $PP$ box, and  in $PB,BB$ only the terms which correspond to configurations that have a $k$-vector in common.
    \item ``$\tau$-threshold'': beyond the full $PP$ box, only include the terms that are higher than a certain threshold $\tau$ in the reduced covariance matrix. 
    \item ``$(\tau)$Common-$k$'': Apply the $\tau$ threshold, as stated above, to the ``Common-$k$'' case.
\end{itemize}
Several additional approximations were initially considered, including  the modifications to the ``Common-$k$'' approach restricting the $BB$ box to only include the off-diagonal terms corresponding to the same triangle $(k_1,k_2,k_3)$, and also the case with all coefficients for $PP$ and $PB$ and with a diagonal $BB$ box. These approximations yielded matrices that were not positive-definite and hence were discarded.  Thresholds over $\sim0.04$ in the $\tau$-threshold approximation are also not positive definite for our present analysis. We adopt a  threshold of 3\%: the resulting  matrix offers a balance between being  as sparse as possible  while  maintaining similarity with the full covariance matrix. Note that there may be a connection  between this value and the findings of \cite{hou2022measurement}, where a level of noise of $\sim2\%$ is reported in the covariance matrices of the \textsc{Patchy} simulations.

\begin{figure}[htb]
\centering 
\includegraphics[ width = \textwidth ]
{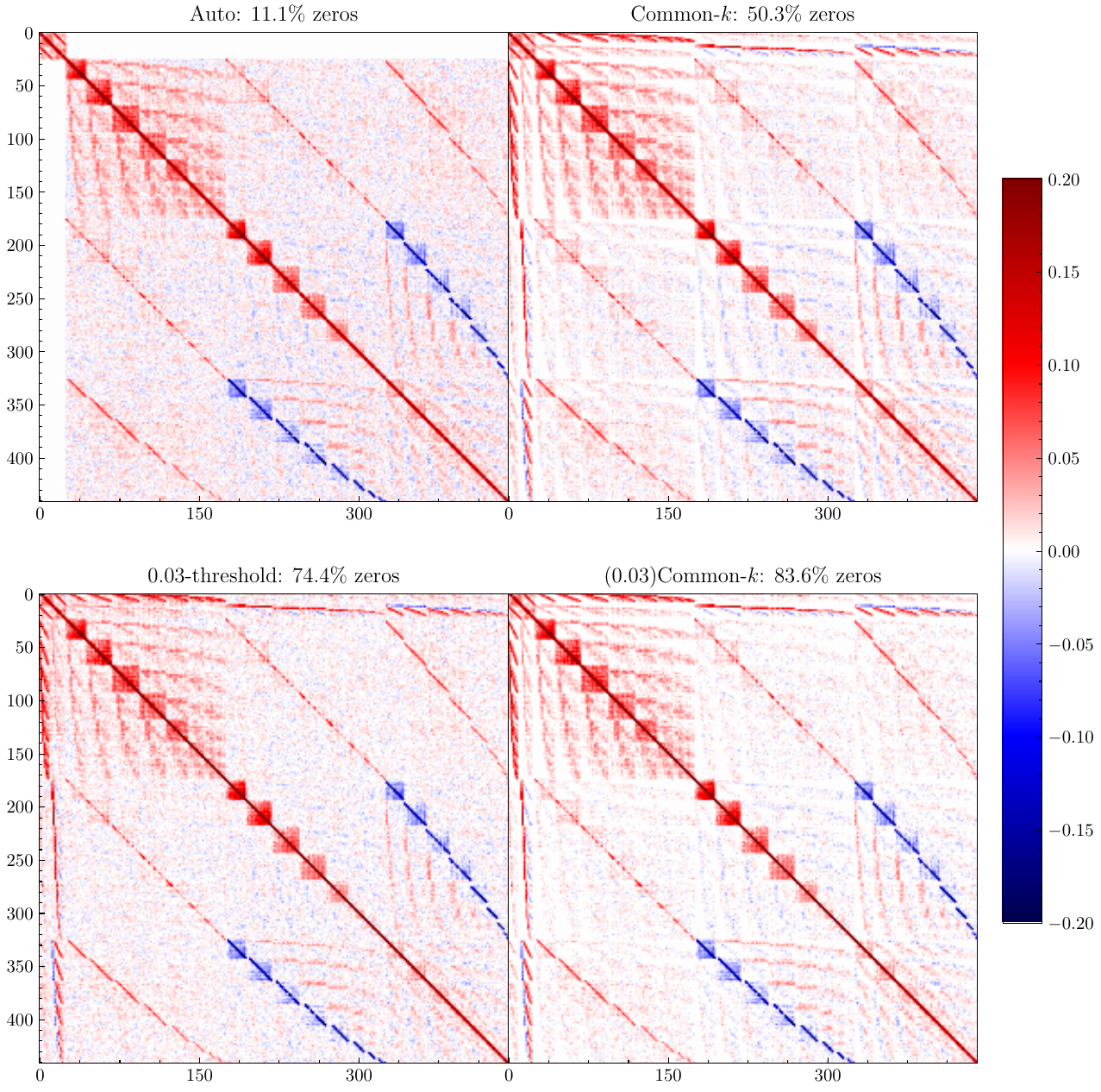}

\caption{ 
Reduced covariance matrices for each of the intermediate approximations to the covariance matrix. We show as well the sparsity of the matrix, represented by the percentage of coefficients that are set to zero in each case. While the Common-$k$ approximation is the one that is closest to the full case while not underestimating the errors (see Figure \ref{fig: eig_approx}), it still only has half of its coefficients set to zero. The figure has an apparent curvature, caused by removing the duplicate isosceles configurations in the bispectrum data-vector, as explained in Section \ref{sec: sims}.}
\label{fig: cov_approxs_red}
\end{figure}

\begin{figure}[htb]
\centering 
\includegraphics[ width = \textwidth ]
{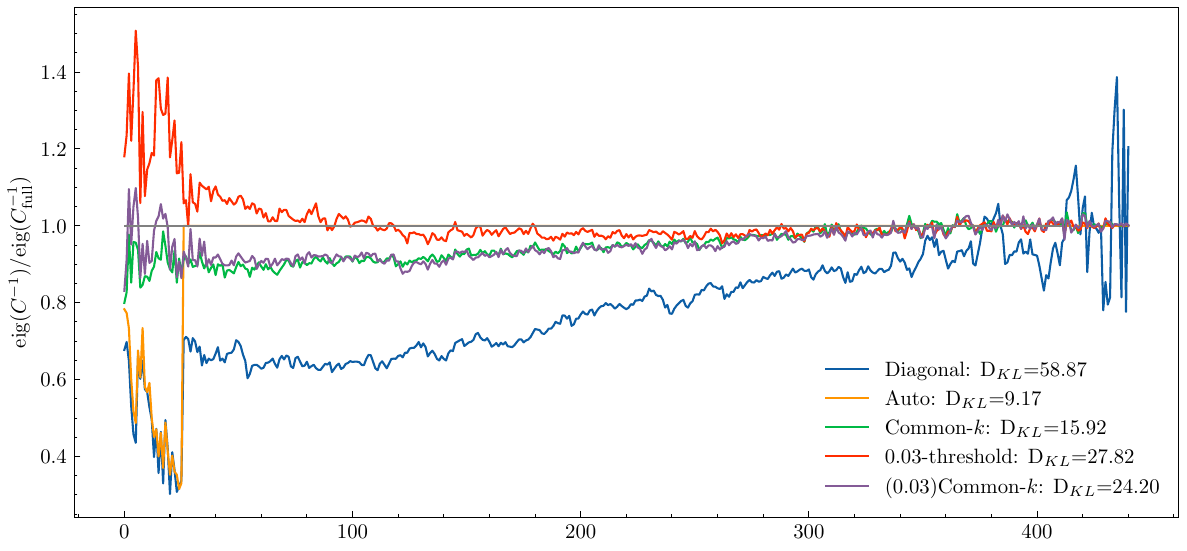}

\caption{ Ratios of the sorted (decreasing order) eigenvalues of the inverse covariance matrix $C^{-1}$ by the inverse full covariance matrix $C^{-1}_{\rm full}$ for all the approximations considered. The eigenvalues of the inverse covariance act as a proxy for the information contained in their direction: the bigger the eigenvalue the more information it potentially contains. The product of the eigenvalues, the determinant, is informative of the size of the recovered error-bars in the derived parameters, so a consistent ratio under unity (as is the case for the diagonal and ``Common-$k$'' approaches) typically results in larger error-bars. The Kullback-Leibler divergence---defined in Equation \ref{eq: kullback}---of each case with respect to the full covariance estimator is shown in the legend.} 
\label{fig: eig_approx}
\end{figure}

A quantitative metric to estimate the ``closeness'' of each approximation  to the reference (full covariance matrix) is the ratio of the eigenvalues. This is reminiscent of the figure-of-merit for the cosmological parameters which is given by  the square root of the determinant of the  Fisher matrix  for the parameters---the determinant is of course the product of the matrix eigenvalues.  The analogy  arises because the (inverse) Fisher matrix is related to the cosmological parameters errors while the  covariance matrix is  related to the data-vector errors. 
The eigenvalues of the inverse covariance matrix can be seen to  represent the  signal or information content of the data-vector. By ordering the eigenvalues in decreasing order it is easier to visualize where most of the information content is localized. 
As expected, the first eigenvalues are those that directly involve the power spectrum.
Moreover, to compress the ``distance'' of an approximation to the full matrix case we also report the logarithm of the ratio of the product of the eigenvalues and the 
Kullback-Leiber (KL) divergence  (or relative entropy, or information gain) between the two distributions that the matrices represent \cite{shannon1948mathematical,kullback1951information}. Assuming that the distributions are multivariate Gaussians,  for our case (where the two have the same mean)  this quantity reads as \cite{tumminello2007kullback}

\begin{equation}
\label{eq: kullback}
    D_{KL}(C_{\rm full}||C)=\frac{1}{2}\left({\rm tr}(C_{\rm full}C^{-1})-n+\ln\left(\frac{\det C}{\det C_{\rm full}}\right)\right),
\end{equation}
 with $n$ the dimension of the data-vector.\footnote{The KL divergence is not symmetric and here we have explicitly expressed the expected excess surprise from using the distribution given by $C_{\rm approx}$ as a model when the actual distribution is that of $C_{\rm full}$.  If the two distributions are multi-variate Gaussians, as we assume here, and have the same mean  (as is the case here, to the best of our knowledge) the KL divergence  assumes this simplified form.}
Figure \ref{fig: eig_approx}  shows the ratios of the (sorted in decreasing order) eigenvalues of the inverse covariance matrix $C^{-1}$ with respect to the full covariance case $C^{-1}_{\rm full}$ for each of the approximations, including the diagonal case. The legend reports the Kullback-Leibler distance, which is complemented by Table \ref{tab: distance_cov} with other indicators of distance towards the full covariance matrix. All this  serves as  figures of merit, to assess  and rank-order both the closeness to the full covariance matrix and also the expected discrepancy in the recovered error-bars. 

\begin{table}[ht]
\centering
\begin{tabular}{|l||c|c|c|}
\hline
Approximation   &  $D_{KL}$ & $\ln\left(\frac{\det C^{-1}}{\det C^{-1}_{\rm full}}\right)$  & $\left(\frac{\det C^{-1}}{\det C^{-1}_{\rm full}}\right)^{1/n}$\\ \hline \hline
Diagonal     & 58.87  & -117.73    & 0.77    \\ 
Auto   & 9.17 & -18.32   & 0.96      \\ 
Common-$k$& 15.92   & -23.33     & 0.95        \\ 
0.03-threshold  & 27.82  & 4.49    & 1.01      \\ 
(0.03)Common-$k$ & 24.20  & -21.66  & 0.95        \\ \hline

\end{tabular}
\caption{Different metrics assessing the distance of the proposed approximations to the full covariance matrix. The Kullback-Leibler distance, $D_{KL}$, is as defined in Equation \ref{eq: kullback}, with one of its components, the logarithm of the ratio of determinants, shown in the second column. The last column quantifies the geometric mean of the ordered ratios of eigenvalues, which is the $n$th root of the ratio of the determinants, with $n$ being the dimension of the data-vector. The diagonal covariance matrix performs significantly worse than the other approximations, while we do not see a gain in the Auto approximation---given that it does not reduce significantly the number of non-zero elements of the covariance matrix (see Figure \ref{fig: cov_approxs_red}). Among the remaining approximations, the 0.03-threshold approach seems to artificially obtain more information, as also seen in Figure \ref{fig: eig_approx}, while we find the Common-$k$ and (0.03)Common-$k$ approximations to strike a good balance between closeness to the full covariance and proportion of zero elements (see Figure \ref{fig: cov_approxs_red}).  }
\label{tab: distance_cov}
\end{table}

The diagonal approximation consistently underestimates the eigenvalues of the inverse matrix. This underestimation is indicative that there is less information that can be captured, which propagates in the error-bars in the parameters, as observed in Section \ref{sec: results} and further discussed below  in Section \ref{sec: theory}. Additionally, its deviation from the full covariance case---as quantified by means of the Kullback-Leibler divergence (Equation \ref{eq: kullback})---is significantly higher than all remaining approximations.

The ``Auto'' approximation, by setting to zero the $PB$ boxes, only changes the first $\sim 25$ eigenvalues, which are those that  change most drastically when including the $PB$ boxes. This indicates that the $PB$ cross-correlation is important and should not be neglected.  Moreover, this approximation does not yield a sparse matrix (only 11\% of the terms are zero) or a matrix that offers any speed-up or computational advantage  compared to the full.

The ``Common-$k$'' approximation has  all eigenvalues much closer to the full covariance case while having $\sim50\%$ of the terms being zero. Therefore, this or a similar approach (together with suitable sparse matrix covariance methods such as \cite{cai2011adaptive,cai2012optimal}) may be advantageous if it can ease the requirements on the number of simulations needed to estimate the covariance.

The 0.03-threshold features a consistent overestimate of the largest eigenvalues of $C^{-1}$, especially for the values concerning the power spectrum. Thus, this approximation on its own  artificially increases, for some eigenvalues,  the information content of the data-vector.

Finally, the (0.03)Common-$k$ approach is a combination of 0.03-threshold and Common-$k$, which yields a covariance matrix whose eigenvalues are closer to the full case than the 0.03-threshold case, while having significantly more sparsity in the matrix (84\%  of the elements are zero, in the case presented in Figure \ref{fig: cov_approxs_red}). Using this approximation in practice would consist of first estimating the Common-$k$ terms---be it fully estimated from simulations or via theoretical model calibrated to simulations e.g., \textit{à la} \cite{Gualdi:2020aniso,fumagalli2022fitting}---to then establish an appropriate threshold for the set-up at hand, keeping in mind that the threshold should maintain the resulting matrix to be positive-definite. Provided that the choice of threshold is conservative, most of the removed coefficients are likely to be noise or noise-dominated, making it a valid approximation of the covariance matrix as a sparse matrix.

\begin{figure}[htb]
\centering 
\includegraphics[ width = \textwidth ]
{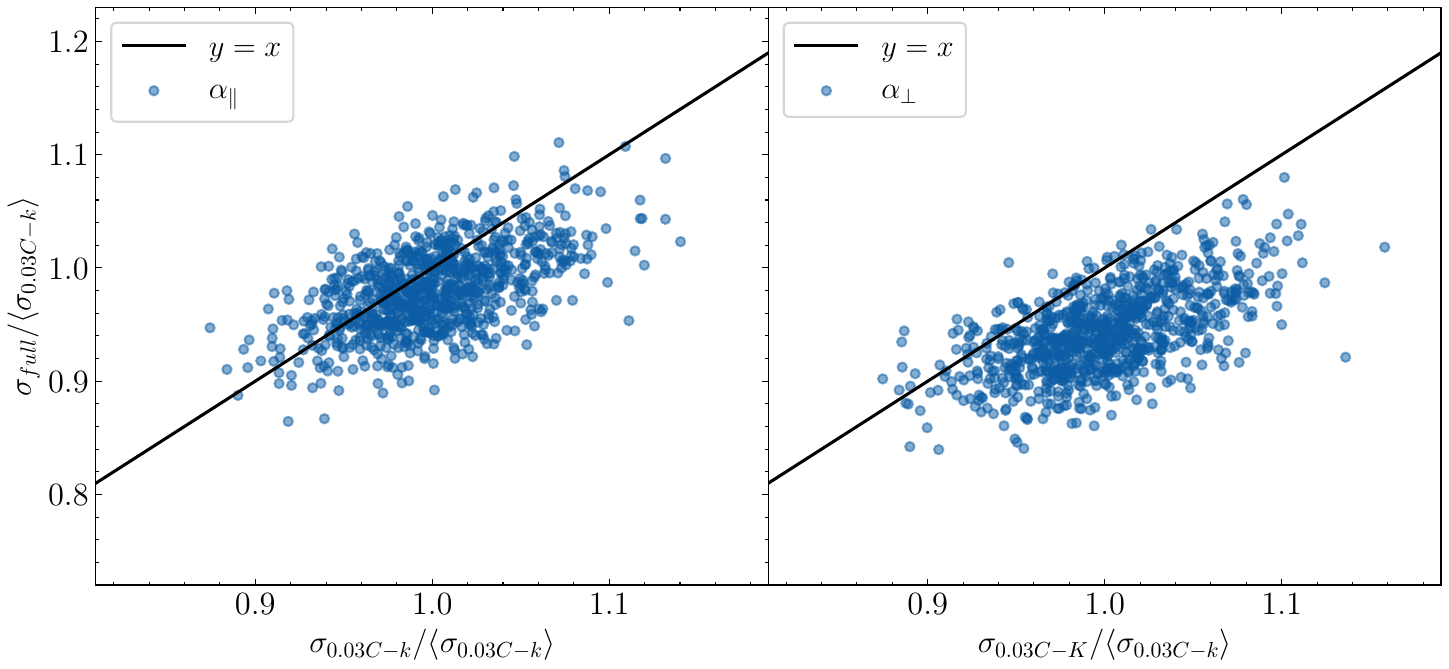}

\caption{Analogous to Figure~\ref{fig: errors_corr_alpha}, but for the (0.03)Common-$k$ vs the full cases.
Compared to Figure~\ref{fig: errors_corr_alpha} the  points are scattered closer to the equality marked by the $y=x$ line indicating that the $\tau$Common-$k$ approximation, in the idealized case of periodic box and with the value of $\tau$ suitably chosen according to the \textsc{Patchy} mocks, yields error-bars only marginally larger than the full covariance matrix There is no appreciable difference of central values for $\alpha_\parallel$, and only in  $\alpha_\perp$ case there is a small mis-estimate $\sigma_{0.03C-k}\sim 1.05\sigma_{full}$ .}
\label{fig: errors_corr_03commk_alpha}
\end{figure}

\begin{figure}[htb]
\centering 
\includegraphics[ width = 0.49\textwidth ]
{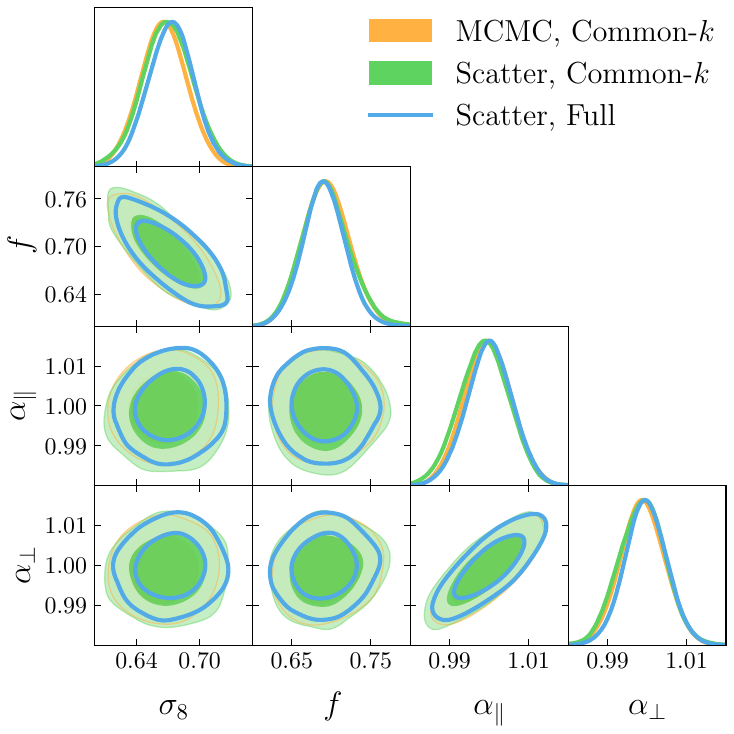}
\includegraphics[ width = 0.49\textwidth ]
{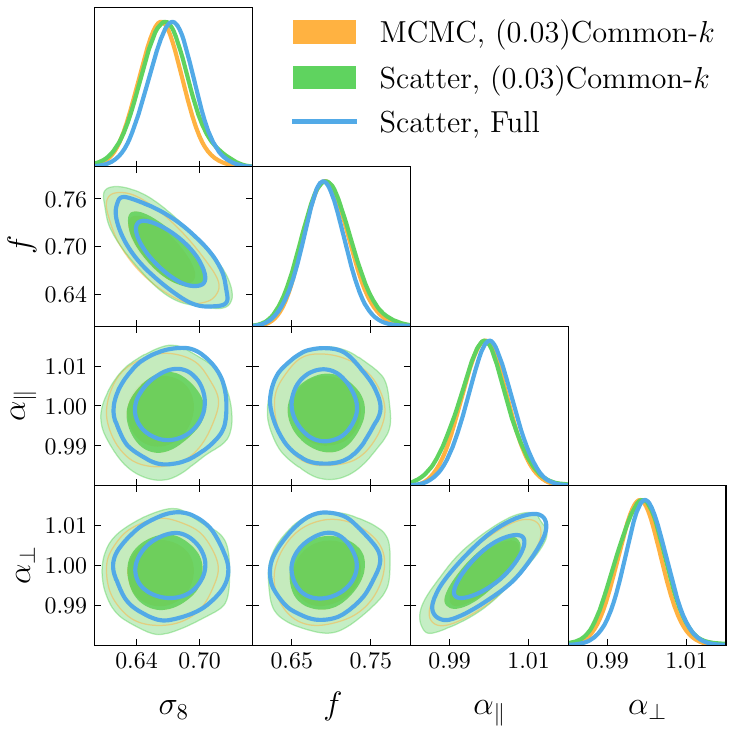}

\caption{Analogous to Figure~\ref{fig: mcmc_vs_scatter}, but focusing on the performance of respectively the ``Common-$k$'' and ``(0.03)Common-$k$'' approximations, compared to the full covariance case. Contrasting with the behaviour of the diagonal approximation displayed in Figure~\ref{fig: mcmc_vs_scatter}, the proposed approximations do not exhibit a significant increase in errors with respect to the full covariance, while maintaining good coverage properties.}
\label{fig: Patchy_03commkcov_1000_pl}
\end{figure}

We repeat the tests carried out in Section \ref{sec: results} for the ``Common-$k$'' and ``(0.03)Common-$k$'' approximations for 1000 \textsc{Patchy} realizations, finding a remarkable agreement with the full covariance matrix in both cases (see Figure \ref{fig: errors_corr_03commk_alpha}, where we report the correlation of errors between the full and ``(0.03)Common-$k$'' covariances). Both approaches result in the \textit{rms} of the $z_\theta$ statistic defined in Equation \ref{eq: def_z} being equal or below 1 ($\sim 1$ -- $0.93$, comparable with the full case of Figures \ref{fig: histz_alpha} and \ref{fig: histz_fs8}) for the parameters $\theta=\{\sigma_8,f,\alpha_\parallel,\alpha_\bot\}$, which is a signal that neither approximation underestimates the error-bars on the key cosmological parameters. We finally show the MCMC constraints and coverage properties for the ``Common-$k$'' and ``(0.03)Common-$k$'' approximations in Figure \ref{fig: Patchy_03commkcov_1000_pl}.

 This figure indicates that the ``Common-$k$''  or the ``$\tau$Common-$k$'' approximations (for a suitable choice of $\tau$  and in particular one that leaves the matrix positive definite) are useful approximations that, in a cosmological parameter inference setting, could provide results virtually indistinguishable from using the full covariance matrix.

\section{Understanding and modeling the  effect of the off-diagonal terms}
\label{sec: theory}

Before we conclude, we attempt to provide an analytic description of the effects illustrated above. 
Although  the content of this section is well-known standard material, having it here helps the reader to put it in context and to better understand the results of the other sections of this paper.

A covariance matrix can always be decomposed in the sum of a (block) diagonal part $\mathbb{D}$ and an off-diagonal part $\epsilon$ so that
\begin{equation}
\mathbb{C}=\mathbb{D}+\mathbb{\epsilon}
\end{equation}
For example, $\mathbb{C}$ could be the full power spectrum plus bispectrum multipoles covariance matrix. See for example \cite{Gualdi:2020aniso} for an analytic expression for the covariance terms and schematic  decomposition in blocks. The effect of the off-diagonal terms within the  power spectrum block of the covariance has been extensively explored in the literature and does not concern us here. We focus on the off-diagonal pure bispectrum terms and  power spectrum-bispectrum terms; the power spectrum  contribution can thus be considered as a block in a block diagonal-dominated matrix.

The magnitude of the $\epsilon_{ij}$ terms compared to the diagonal contribution to the covariance of the bispectrum has been studied in Appendix A of \cite{novell2023geofpt}. There it is  shown that $r_{ij}=\epsilon_{ij}/\sqrt{\mathbb{D}_i\mathbb{D}_j}$  when averaged over off-diagonal blocks involving  the bispectrum,  takes values  up to 0.01, but when not averaged it is as high as 0.6.

If it could be assumed that $\mathbb{C}$ was diagonal-dominant (that is $\epsilon _{ij}\ll \sqrt{\mathbb{D}_i\mathbb{D}_j}$), then  a possible approximation  for the  inverse of $\mathbb{C}$ could be \cite{sherman1950adjustment,woodbury1950inverting}
\begin{equation}
\label{eq: diag_dominant}
 \mathbb{C}^{-1}\simeq \mathbb{D}^{-1}-\mathbb{D}^{-1}\epsilon \mathbb{D}^{-1}\equiv \mathbb{D}^{-1}-\mathbb{M}  \equiv \mathcal{C}^{-1}
\end{equation}
 where the last equalities define the matrices $\mathbb{M}$ and  $\mathcal{C}^{-1}$.  For  the  covariance matrix involving the bispectrum  we find that for the \textsc{Patchy} mocks the median residual between $\mathbb{C}^{-1}$ and $\mathcal{C}^{-1}$ is -0.62, which we illustrate in Figure \ref{fig: discrepancy_invC} in Appendix \ref{app: cov_graphical}. Hence, the diagonal-dominant approximation is completely invalid for the data-vector of interest, which already signals the importance of the off-diagonal terms of the covariance matrix. We, therefore, define the off-diagonal contribution to the covariance as the matrix $\mathbb{O}$ fulfilling $\mathbb{C}^{-1}\equiv  \mathbb{D}^{-1}-\mathbb{O}$.

Note that Equation~\ref{eq: diag_dominant} is more general than stated above since $\mathbb{D}$ does not need to be diagonal for the approximation to be valid  provided that the   matrix $\epsilon$ is a perturbation to the matrix $\mathbb{D}$. For example, the matrix $\mathbb{D}$ could be the thresholded  covariance matrix, and $\epsilon=\mathbb{C}-\mathbb{D}$.\footnote{We have checked that for the thresholds considered in this specific application this approximation is not good enough.}

The effect of the off-diagonal components on the $\chi^2$ and on the response of the $\chi^2$ to a change of parameters can be modeled as follows. Let us define the vector ${\bf u}={\bf d}-{\bf t}$ where ${\bf d}$ denotes the data (measurements) vector and ${\bf t}$ the theoretical model. Then 
\begin{equation}
    \chi^2_{Diag}=\sum_i { u}_i^2\mathbb{D}_{ii}^{-1}=\sum_i \frac{{ u}_i^2}{\mathbb{D}_{ii}}
\end{equation}
\begin{equation}
\chi^2_{ND}=- \sum_{i, j} { u}_i \mathbb{O}_{ij} { u}_j
\label{eq:chosqnd}
\end{equation}
and of course $\chi^2=\chi^2_{Diag}+ \chi^2_{ND}$. Equation \ref{eq:chosqnd} (and the following equations involving $\mathbb{O}$),  if $r_{ij}\ll1$ for $i\neq j$,  could be approximated  by substituting $\mathbb{M}$ for $\mathbb{O}$. In the applications considered here, this approximation is not sufficient.

For the \textsc{Patchy} analysis from Section~\ref{sec: results}, we find the typical contribution of the off-diagonal part of the covariance to $\chi^2$ (evaluated at the best-fit cosmological parameters) to be  of $|\chi^2_{ND}|\sim0.04\chi^2$. This is a small proportion, but it should be noted that the importance and role of the off-diagonal terms in the covariance do not necessarily correspond to their effect on the $\chi^2$, but rather to its response, $\Delta \chi^2$.

The response of the likelihood and the $\chi^2$ to a change in cosmological parameters from, say, the best-fit values, that induces a change $\Delta t$ on the theory model from the best-fit theory model, can be written as:
\begin{equation}
   \Delta  \chi^2_{Diag}=-2\sum_i { u}_i\mathbb{D}_{ii}^{-1}\Delta t_i=-2 \sum_i \frac{u_i\Delta t_i}{\mathbb{D}_{ii}}
\end{equation}
\begin{equation}
\Delta \chi^2_{ND}=-2 \sum_{i, j} {u}_i \mathbb{O}_{ij} \Delta t_j
\label{eq:Dchosqnd}
\end{equation}
 where again  $\Delta \chi^2=\Delta \chi^2_{Diag}+ \Delta \chi^2_{ND}$.
 Figure~\ref{fig: errors_corr_alpha} shows that there is a weak correlation between  $\Delta \chi^2_{Diag}$ and $\Delta \chi^2$, and that  $\Delta \chi^2_{ND}$ is not negligible compared to  $\Delta \chi^2_{Diag}$. 
 
For an idealized case where ${\bf d}$ is noiseless and ${\bf t}$ is a perfect description of ${\bf d}$,  for example when doing Fisher matrices forecasts, ${\bf u}={\bf d}-{\bf t}=0$: the size of the errors  can be estimated from the second derivative of the log-likelihood (see below). In any practical realization  ${\bf u}\ne 0$.

Note that in the case of the bispectrum covariance, the  relative importance of ($\Delta$)$\chi^2_{ND}$ and  ($\Delta$)$\chi^2_{Diag}$ to ($\Delta$)$\chi^2$ do not depend on the  survey volume used to rescale the covariance. Hence (relative) results obtained for $1\, {\rm Gpc}^3$ (which is much smaller than the volume of ongoing surveys)   also hold for stage IV dark energy experiments with volumes $\sim  50\,{\rm Gpc}^3$. As shown even visually for example  in  \cite{Gualdi:2020aniso}, the matrix elements of $\mathbb{D}$ and $\epsilon$ and therefore  the matrix elements of $\mathbb{D}^{-1}$  and  $\mathbb{O}$ both  scale with the survey volume coherently, provided everything else including the choice of binning and the  effect of the survey window remain unaltered. However, different survey volumes  or different survey geometries  can enable  different binning choices in $k$-space and thus different correction factors $A_P$ and $A_B$ (using  nomenclature from \cite{Gualdi:2020aniso}) and different scaling in the diagonal with respect to the off-diagonal elements. The effect of different shot noise levels might also  rescale the diagonal and non-diagonal contributions to ($\Delta$)$\chi^2$ differently. Moreover, the realization of ${\bf d}$ (and  ${\bf u}$) will be more or less noisy depending on the volume of the realization.

The decomposition in terms of a matrix and a perturbation can help us understand also the behaviour of the eigenvalues.
Let us consider the sum of matrices $\mathbb{C}=\mathbb{D}$+$\mathbb{O}$, with their eigenvalues being respectively: $c_1\ge..\ge c_n$; $\mathbb{D}$: $d_1 \ge ...\ge d_n$;  $\mathbb{O}$:   $o_1 \ge ...\ge o_n$.

Weyl's inequality states that:
$d_i+o_n\le c_i\le d_i+o_1$. This inequality is valid for any sum of symmetric matrices, i.e., holds even if $\mathbb{D}$ is  not diagonal.
If  $\mathbb{O}$ is a perturbation of  $\mathbb{D}$ of order $\epsilon$ then 
$|c_i-d_i|\le \epsilon$. In fact, if we  identify $\mathbb{D}$  with the thresholded matrix $\mathbb{T}$ and  $\mathbb{O}$ with $\mathbb{C}-\mathbb{T}=\epsilon$, the thresholded matrix  $\mathbb{T}$ will have each eigenvalue bounded below (resp. above) by the sum of the eigenvalue of $\mathbb{C}$ and the minimum (resp. maximum) eigenvalue of $\mathbb{O}$. In this case,  the matrix $\mathbb{O}$ has approximately random values on the coefficients that are not set to zero, and then the minimum and maximum eigenvalues will be respectively negative and positive. Therefore, the matrix $\mathbb{T}=\mathbb{C}+\mathbb{O}$ can have eigenvalues lower than $\mathbb{C}$.

\subsection{Relation to Fisher forecasts}

We can now see  in a transparent way the effect of the off-diagonal covariance terms in a Fisher-forecast analysis. 
In this approach to forecasts, the Fisher information matrix $\mathbb{F}$ yields statistical error estimates on the model's parameters $\{\theta_\alpha,\theta_\beta...\}$. The Greek indices hereafter will run over   the model's parameters.
Given ${\cal L}=-\ln L$ where $L$ denotes the likelihood, which under the Gaussian approximation ${\cal L}=\chi^2/2$,
\begin{equation}
    \mathbb{F}_{\alpha \beta}=\left\langle\frac{\partial ^2{\cal L}}{\partial\theta_\alpha \partial \theta_\beta} \right\rangle =\sum_{ij}\partial_\alpha t_i \mathbb{C}^{-1}_{ij}\partial_\beta t_j= \sum_{i}\partial_\alpha t_i \mathbb{D}^{-1}_{ii}\partial_\beta t_i-\sum_{ij}\partial_\alpha t_i \mathbb{O}_{ij}\partial_\beta t_j\equiv \mathbb{F}'_{\alpha\beta}-{\cal F}_{\alpha \beta}
\end{equation}
where linear dependence of the theory on the parameters is assumed and we have decomposed the Fisher matrix in a diagonal part  $\mathbb{D}$ and a purely off-diagonal contribution $\mathbb{O}$.  In the above equation $\mathbb{O}$ can be replaced by  $\mathbb{M}$ only when the off-diagonal contributions to $\mathbb{C}$ are small enough.
The marginal error on a given parameter is thus $\sigma_\alpha=(\mathbb{F}^{-1})^{1/2}_{\alpha \alpha}$.
\begin{equation}
\label{eq: fisher_sigma}
\sigma^2_{\alpha}=[(\mathbb{F}\,'-{\cal F})^{-1})]_{\alpha \alpha}= [\mathbb{F}\,'^{-1}]_{\alpha \alpha}+ [\mathbb{F}\,'^{-1} {\cal F}\, (\mathbb{F}\,'')^{-1}]_{\alpha\alpha}\equiv \sigma^2_{Diag, \alpha}+\Delta\sigma^2_{ND\alpha}
\end{equation}
where  $ \mathbb{F}\,''=(\mathbb{F}\,'-{\cal F})$ (a recursive expression) and  $\Delta\sigma^2_{ND\alpha}$ is the correction to the diagonal errors induced by the off-diagonal terms. If $\mathbb{F}\,''$ is approximated by $\mathbb{F}\,'$
 one is assuming that the conditional errors still coincide with the marginalized errors, i.e., $\Delta\sigma^2_{ND}=0$ (as it happens for a diagonal Fisher matrix). The resulting (Fisher) error ellipses will display approximately correct degeneracies, but the values for the marginalized errors will be underestimated.

\begin{figure}[htb]
\centering 
\includegraphics[ width = \textwidth ]
{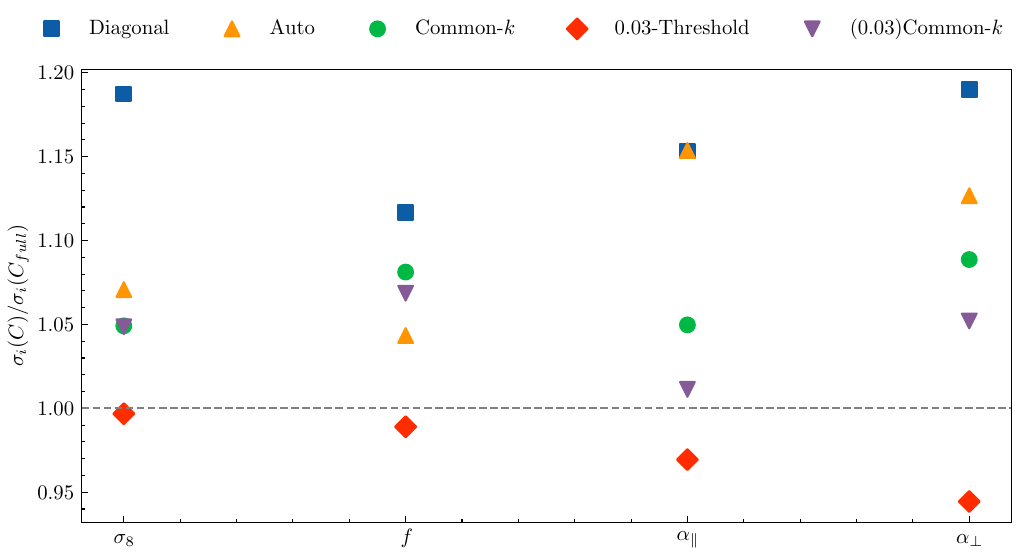}

\caption{Fisher forecasts for the parameters of interest $\{\sigma_8,f,\alpha_\parallel,\alpha_\bot\}$ for all the  approximations considered in this work. These are shown, for each of the parameters,  as the ratio between forecasted  errors using the covariance approximation over the  errors obtained with the full covariance.  All but the ``0.03-thresholded'' case result in higher values of $\sigma$.
The diagonal case gives the overall largest errors (as expected). The ``Auto'' case  is shown  for reference as it ignores the $PB$ terms but includes all the off-diagonal $BB$ terms and thus does not offer any  significant advantages or speed up in its evaluation.
}
\label{fig: covmat_fisher}
\end{figure}

In our specific case, we  perform Fisher forecasts for the cosmological parameters of interest  with  the full matrix and various approximations to it. It is  then possible to compare $\sigma^2_{Diag}$ with $\sigma^2$ for the parameters of interest, finding that the Fisher forecasts for the diagonal covariance matrix are 10-20\% larger than those of the full covariance matrix---as illustrated by Figure \ref{fig: covmat_fisher}.

We find that, for the analysis set-up featuring the \textsc{Patchy} mocks, the median element of ${\cal F}/\mathbb{F}\,'$ is 0.43, so the approximation $\mathbb{F}\,''\simeq \mathbb{F}\,'$ will introduce a sizeable error in the forecasted marginalized errors.

In a realistic scenario, the vector \textbf{u} is non-zero, and the estimator will only be efficient if the full covariance is used, which provides the optimal weighting of the data-vector. Ignoring the off-diagonal terms effectively weights the data-vector in a sub-optimal way, yielding higher uncertainty on the recovered parameter values. While the case using the full covariance approximates the Cramer-Rao bound, the case using the diagonal covariance approximation loses information, resulting in a less efficient maximum likelihood estimator with (possibly underestimated) bigger error-bars. 

\section{Discussion and Conclusions}
\label{sec: concl}

The inclusion of the bispectrum multipoles in cosmological parameters estimations from the large scale structure improves significantly cosmological constraints (see e.g. \cite{novell2023geofpt} for a quantitative assessment). But this information gain  comes at a cost. 
Besides the additional effort needed to model the bispectrum signal, in the joint likelihood analysis of the redshift space multipoles of the power spectrum and bispectrum, the evaluation of the full covariance matrix can be challenging.  The inclusion of the bispectrum multipoles increases dramatically the data-vector size compared to a power spectrum-only analysis. This increases the requirement on the number of mock simulations needed to evaluate reliably the covariance matrix from simulations.

All these challenges could be significantly eased if suitable approximations of the covariance matrix are found. 

We have shown that approximating the full covariance matrix  blocks that include the bispectrum with a purely diagonal matrix means using an unbiased but sub-optimal estimator. This causes the true errors on the recovered cosmological parameters to be not optimal, and the inferred error-bars to be underestimated, with the constraints not having ``coverage''.
The standard likelihood analysis with the full covariance matrix, if anything, slightly overestimates the errors estimated from the scatter among recovered parameters from  many realizations. However, this is likely  an apparent error overestimation due to the fact that the scatter among simulations is done with an approximate covariance matrix assuming fully Gaussian statistics  and does not fully propagate the uncertainty in the covariance estimation itself \cite{Sellentin:2015waz}; in fact, the magnitude of the apparent overestimation is correlated with the square-root of the  Hartlap factor.

We have quantified these effects on a suite of simulated boxes of dark matter and more realistic dark matter tracers, although the effects reported are expected to be even larger in more realistic set-ups such as with survey window and selection functions.
The covariance matrix estimated from simulations is {\it dense}: there are many small off-diagonal terms but none of them is exactly zero.
While ``thou shall not ignore the bispectrum  covariance off-diagonal terms'',
other approximations may be advantageous. 
In particular, approximations that make the matrix sparse instead (where many off-diagonal terms are zero instead of being very small but non-zero) can be particularly interesting.  The  requirements on the number of simulations, for example, may be, potentially, relaxed. 

Also,   analytic expressions (which can
be long and cumbersome to evaluate for all terms) can be evaluated only for the few terms that really matter and more easily calibrated on simulations  (following for example the prescription of \cite{Gualdi:2020aniso}). The best approximations  for the block involving the bispectrum  we recommend using  are either %
considering to be non-zero only the terms that have a $k$ mode in common, or applying a suitable thresholding operation to the matrix. The first approach, ``Common-$k$'', keeps $\sim45\%$ of the bispectrum auto-covariance terms, while the second approach, ``$(\tau)$Common-$k$'', keeps between 15 and 20\% of the bispectrum auto-covariance terms, depending on the $k$-binning. Such approximations produce results  for cosmological inference virtually indistinguishable from, and with the same coverage properties, as those obtained with the full covariance matrix.

We envision that  these findings  will be useful for the joint redshift space  power spectrum and bispectrum analyses from forthcoming surveys, once real-world effects such as window and selection functions and systematics weights are included in the modeling.

\section*{Acknowledgements}
SNM acknowledges funding from the official doctoral program of the University of Barcelona for the development of a research project under the PREDOCS-UB grant.
HGM acknowledges support through the program Ram\'on y Cajal (RYC-2021-034104) of the Spanish Ministry of Science and Innovation. 
LV and HGM acknowledge the support of the European Union’s Horizon 2020 research and innovation program ERC (BePreSySe, grant agreement 725327).

Funding for this work was partially provided by the Spanish MINECO under project PGC2018-098866-B-I00MCIN/AEI/10.13039/501100011033 y FEDER ``Una manera de hacer Europa'', and the ``Center of Excellence Maria de Maeztu 2020-2023'' award to the ICCUB (CEX2019-000918-M funded by MCIN/AEI/10.13039/501100011033).

This work has made extensive use of the following publicly available codes: \href{https://github.com/serginovell/geo-fpt}{\textsc{GEO-FPT}}, \href{https://emcee.readthedocs.io/en/stable/index.html}{\textsc{Emcee}},\href{https://www.gnu.org/software/gsl/}{GSL}, \href{https://scipy.org/}{\textsc{SciPy}}, \href{https://numpy.org/}{\textsc{NumPy}}, \href{https://getdist.readthedocs.io/en/latest/}{\textsc{GetDist}}, \href{https://matplotlib.org}{\textsc{Matplotlib}}. We are grateful  to the developers who made these codes public.

\appendix
\section{MCMC and simulation scatter error for $\sigma_8$ and $f$}

\begin{figure}[htb]
\centering 
\includegraphics[ width = \textwidth ]
{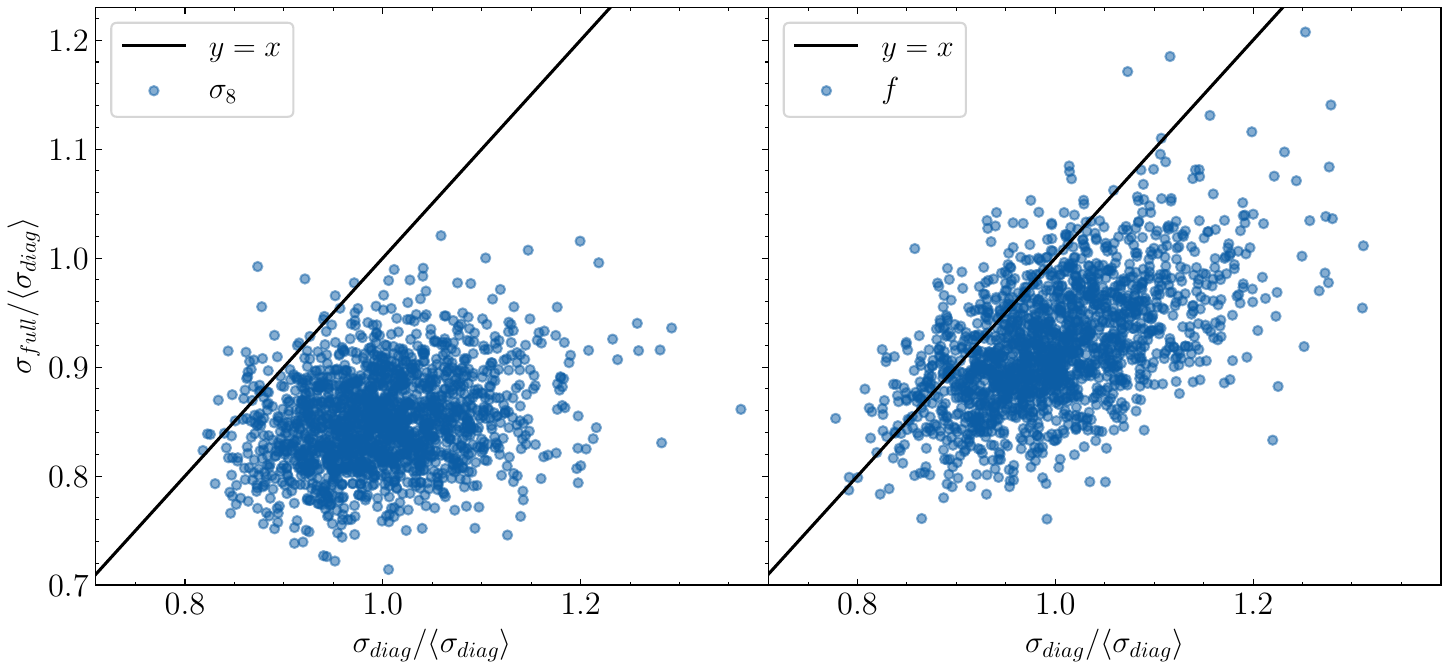}

\caption{Analogous plot to Figure \ref{fig: errors_corr_alpha}, for the case of the parameters $\sigma_8$ and $f$. Similar results are achieved, with the slight variation that the errors on $f$ feature a smaller ratio, $\sigma_{\rm diag}\sim1.1\sigma_{\rm full}$, and that $\sigma_8$ has less correlation between the diagonal and full covariance recovered errors. }
\label{fig: errors_corr_fs8}
\end{figure}

\begin{figure}[htb]
\centering 
\includegraphics[ width = \textwidth ]
{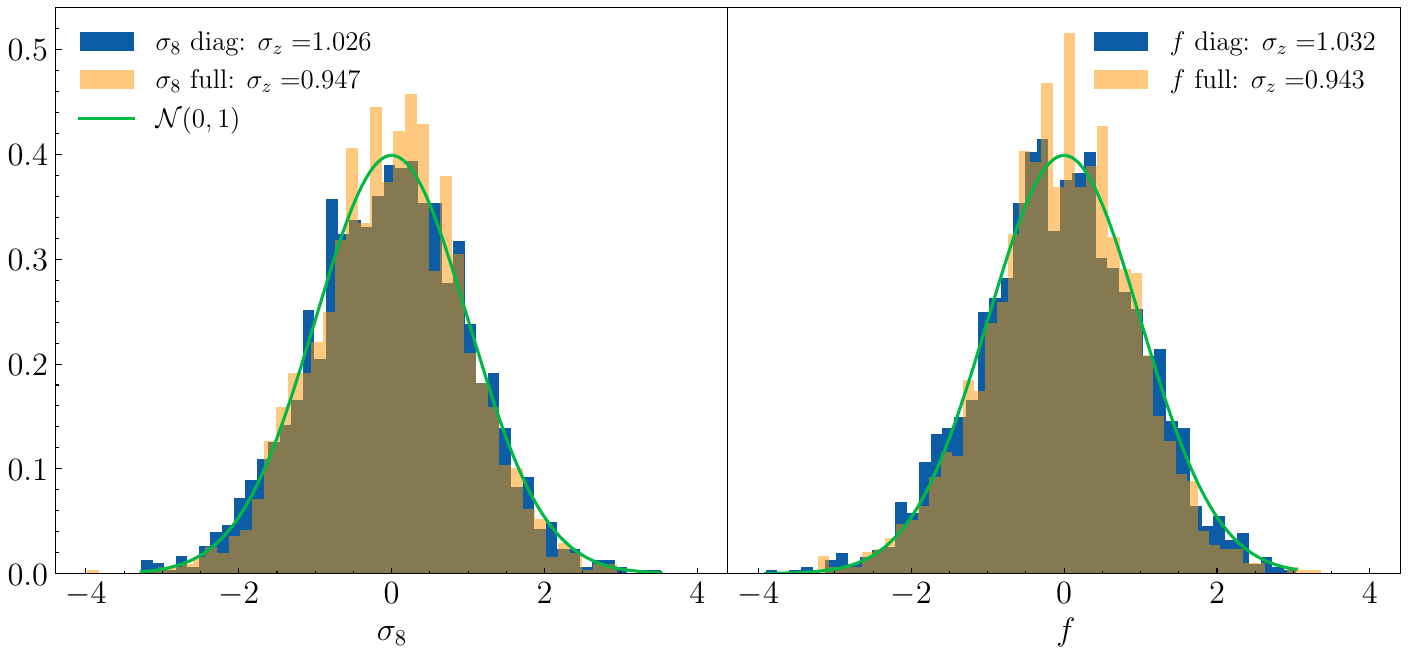}

\caption{Analogous plot to Figure \ref{fig: histz_alpha}, for the parameters $\sigma_8$ and $f$. For these parameters, both the error underestimation (resp. overestimation) for the diagonal (resp. full) covariance is less marked. }
\label{fig: histz_fs8}
\end{figure}

 Figures \ref{fig: errors_corr_fs8} and \ref{fig: histz_fs8} are the counterpart of the Figures in  Section \ref{sec: results} for the remaining parameters of interest, $\sigma_8$ and $f$. A qualitatively similar behaviour as for  the parameters $\alpha_\parallel,\alpha_\bot$ is obtained. The only minor differences (appearing in Figure \ref{fig: errors_corr_fs8}) are that $\sigma_8$ features less correlation between $\sigma_{\rm diag}$ and $\sigma_{\rm full}$ and that the diagonal covariance errors on $f$ are closer to the case with the full covariance---in a factor of $\sim10\%$.   
\label{app: fs8}

\section{Visualisation of the limitations of the diagonal approximation}
\label{app: cov_graphical}
\begin{figure}[htb]
\centering 
\includegraphics[ width = 0.5\textwidth ]
{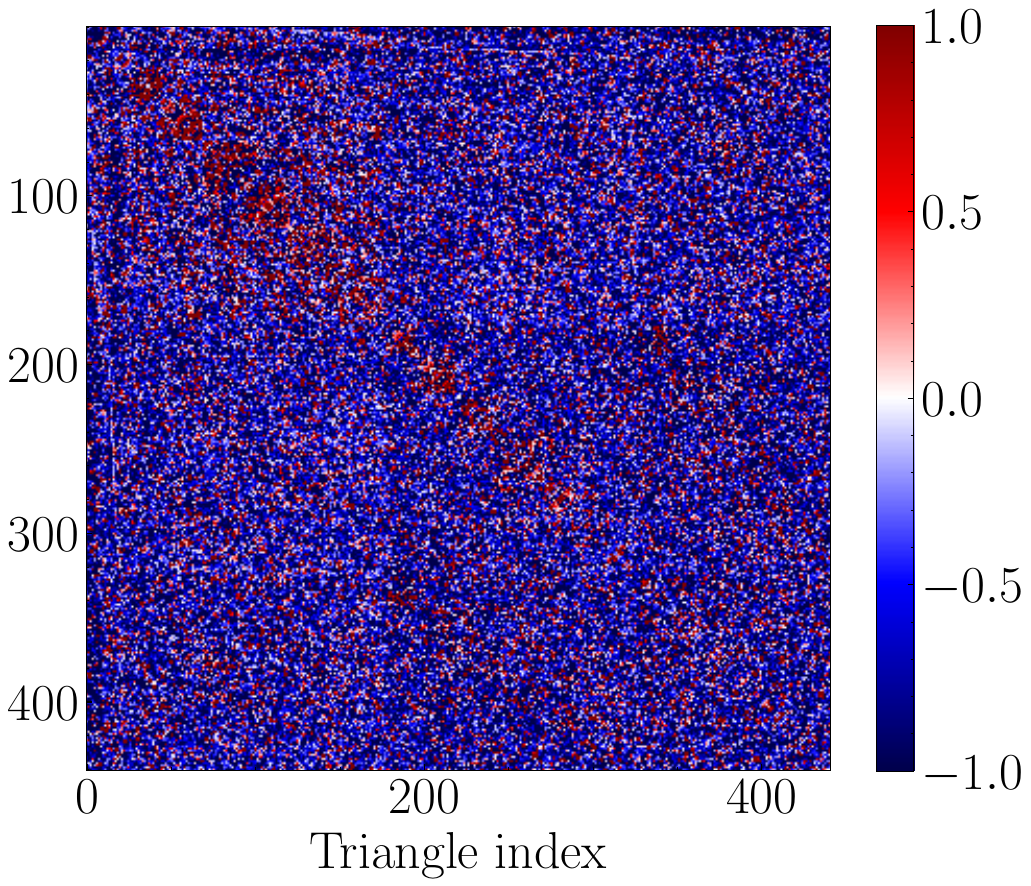}
\includegraphics[ width = 0.485\textwidth ]
{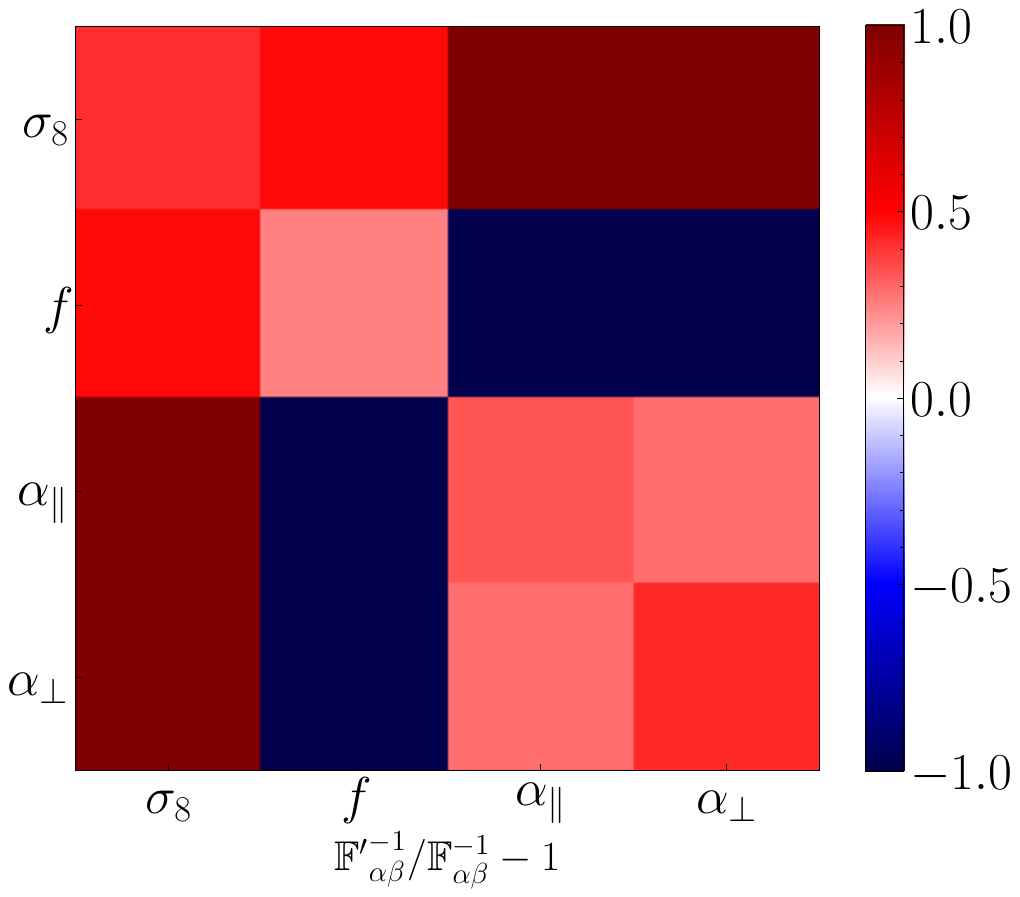}
\caption{Discrepancy in the obtained inverse covariance matrix by assuming that the covariance matrix is diagonal-dominant, $\mathcal{C}^{-1}=\mathbb{D}^{-1}-\mathbb{D}^{-1}\epsilon \mathbb{D}^{-1}$, and the accurate estimation of $C^{-1}$. In the left plot we show the residuals $(C^{-1}-\mathcal{C}^{-1})/C^{-1},$ element-wise. 
The residuals for the inverse Fisher matrix obtained from the diagonal approximation ($\mathbb{F'}$) and from the full covariance matrix ($\mathbb{F}$) are shown in the right plot. In both, it is clearly visible how the diagonal-dominant approximation is invalid for the analysis present in this work, signaling the importance of the off-diagonal terms.  }
\label{fig: discrepancy_invC}
\end{figure}

In Figure \ref{fig: discrepancy_invC} we further  quantify the error introduced by  assuming the covariance matrix to be diagonally dominant. As said in Section \ref{sec: theory}, in such a case, the inverse covariance matrix could be approximated as in Equation \ref{eq: diag_dominant}: $\mathcal{C}^{-1}=\mathbb{D}^{-1}-\mathbb{D}^{-1}\epsilon \mathbb{D}^{-1}$. We show that this would introduce unacceptable biases to the estimation of the inverse covariance matrix, which would propagate to the cosmological analysis. Hence, the off-diagonal terms ought not to be ignored.

A complementary way to observe the effect of assuming a diagonal covariance is shown in the right panel of Figure \ref{fig: discrepancy_invC}. There, the residual between the inverse Fisher matrices when assuming a diagonal and full covariance are displayed, which are notated respectively as $\mathbb{F}$ and $\mathbb{F'}$ following the conventions of Section \ref{sec: theory}. As in Figure \ref{fig: covmat_fisher}, we see how using diagonal covariance results in a forecasted error in the 1D marginalized constraints of $\{\sigma_8,f,\alpha_\parallel,\alpha_\bot\}$ between 10 and 20\% higher than using the baseline full covariance. In this figure we can additionally see that the cross-correlations between parameters will be affected to a higher degree in percentage than the auto-correlations.

\section{Additional comments on the $z$ statistic}
\label{app: z}
Let us further unpack the $z$ statistic defined in Equation \ref{eq: def_z} as
\begin{equation}
    z_{\theta_j}\equiv\frac{\theta_j-\mu_{\theta_j}}{\sigma(\theta_j)}.
\end{equation}

The samples of $z$ can be seen as a 2D array with indices representing the mock realization ($i$) and the parameter ($j$). For each single parameter $j$ and each of the 2000 mock realizations, we conduct parameter inference for each, deriving a best-fit parameter ($\theta^i_j$) and an \textit{estimate} of its error ($\sigma^i(\theta_j)$). It is important to note that using an incorrect covariance in this inference results in an estimator that is unbiased but not optimal, leading to potentially inaccurate error estimates.

The average over 2000 realizations of the best fit parameters has a very little error and can be taken to correspond to the true values $\mu_j$.  So, the scatter of the 2000 best fit values $\theta$ is an estimate of the error which, for a sufficiently large number of realizations, should be very close to the true error---furthermore, by construction it has good coverage properties. This approach is a standard \textit{rms} estimate and does not involve Hartlap or Sellentin-Heavens corrections as no matrix inversion is needed. For 2000 realizations we expect the error on the error to be quite small and at all effects it can be taken as the true error. 

According to the central limit theorem, the distribution of $z_{\theta_j}$ should approach a Gaussian form. The width of this distribution will be unity if $\sigma^i(\theta_j)$ accurately estimates the true error on $\theta_j$, i.e. if it has equivalent coverage to the scatter of $\theta_j^i$ among the 2000 realizations. However, the extent to which the central limit theorem is applicable remains uncertain, and the assumption of a Gaussian likelihood with fixed covariance for bispectrum data is known to be an approximation and might be slightly sub-optimal. Additionally, the Sellentin-Heavens correction can further deviate an initially Gaussian likelihood from Gaussianity.

Our findings in Section \ref{sec: results} indicate that using a diagonal covariance approximation tends to underestimate errors by about 15\%. Conversely, employing the full covariance estimated from simulations, while using the Sellentin-Heavens correction and assuming a Gaussian likelihood for the data vector, may slightly overestimate the Gaussianized errors as determined from the scatter under the central limit theorem.

However, some mismatch in these estimations is anticipated for at least two reasons: firstly, the estimated covariance has an associated error, since this estimate might vary if different sets of mocks are used. This variance is in part why the Sellentin-Heavens correction effectively increases the error estimates. Secondly, the likelihood is not perfectly Gaussian, and matching the \textit{rms} to a $1\sigma$ of a Gaussian distribution may result in a slight discrepancy.

\bibliographystyle{ieeetr}
\bibliography{references}

\end{document}